\documentclass[namedreferences]{SolarPhysics}
\usepackage[optionalrh,solaenum]{spr-sola-addons} 
\usepackage{graphicx}        
\usepackage{color}           
\usepackage{hyperref}             

\newcommand{\etal}{{\it et al.}}

\newcommand{\ang}{{\mathrm \AA}}
\newcommand{\citext}[1]{\citeauthor{#1} (\citeyear{#1})}
\newcommand{\citexts}[2]{\citeauthor{#1} (\citeyear{#1}, \citeyear{#2})}

\newcommand{\aap}{    {\it Astron. Astrophys.}}

\newcommand{\apj}{    {\it Astrophys. J.}}

\newcommand{\nat}{    {\it Nature}}

\newcommand{\pasj}{   {\it Pub. Astron. Soc. Japan}}

\newcommand{\solphys}{{\it Solar Phys.}}

\newcommand{\ssr}{    {\it Space Sci. Rev.}}

\begin{document}

\begin{article}

\begin{opening}

\title{Observations of Unresolved Photospheric Magnetic Fields in Solar Flares Using Fe~{\sc I} and Cr~{\sc I} Lines}

\author{M.~\surname{Gordovskyy}$^1$ \sep 
        V.G.~\surname{Lozitsky}$^2$  
       }
\runningauthor{M. Gordovskyy, V.G. Lozitsky}
\runningtitle{Unresolved Magnetic Fields in Solar Flares}

   \institute{$^1$Jodrell Bank Centre for Astrophysics, University of Manchester, Manchester M13 9PL, UK.
                     email:~\url{mykola.gordovskyy@manchester.ac.uk}  \\
	$^2$Astronomical Observatory, Kyiv National University, Observatorna 3, Kyiv 01053, Ukraine.
             }

\begin{abstract}
The structure of the photospheric magnetic field during solar flares is examined using {\it echelle} spectropolarimetric
observations. The study is based on several Fe~{\sc I} and Cr~I lines observed at locations 
corresponding to brightest H$\alpha$ emission during thermal phase of flares. The analysis is performed by comparing magnetic field values deduced from lines with different magnetic sensitivities, as well as by
examining the fine structure of $I\pm V$ Stokes profiles splitting. It is shown that the field has at least two components, 
with stronger unresolved flux tubes embedded in weaker ambient field. Based on a two-component magnetic field model, we compare observed and synthetic line profiles and show that the field strength in small-scale flux tubes is about $2-3$~kG. Furthermore, we find that the small-scale flux tubes are associated with flare emission, which may have implications for flare phenomenology.
\end{abstract}


\end{opening}


\section{Introduction}
   \label{intro}

\begin{figure}   
\centerline{\includegraphics[width=1.0\textwidth,clip=]{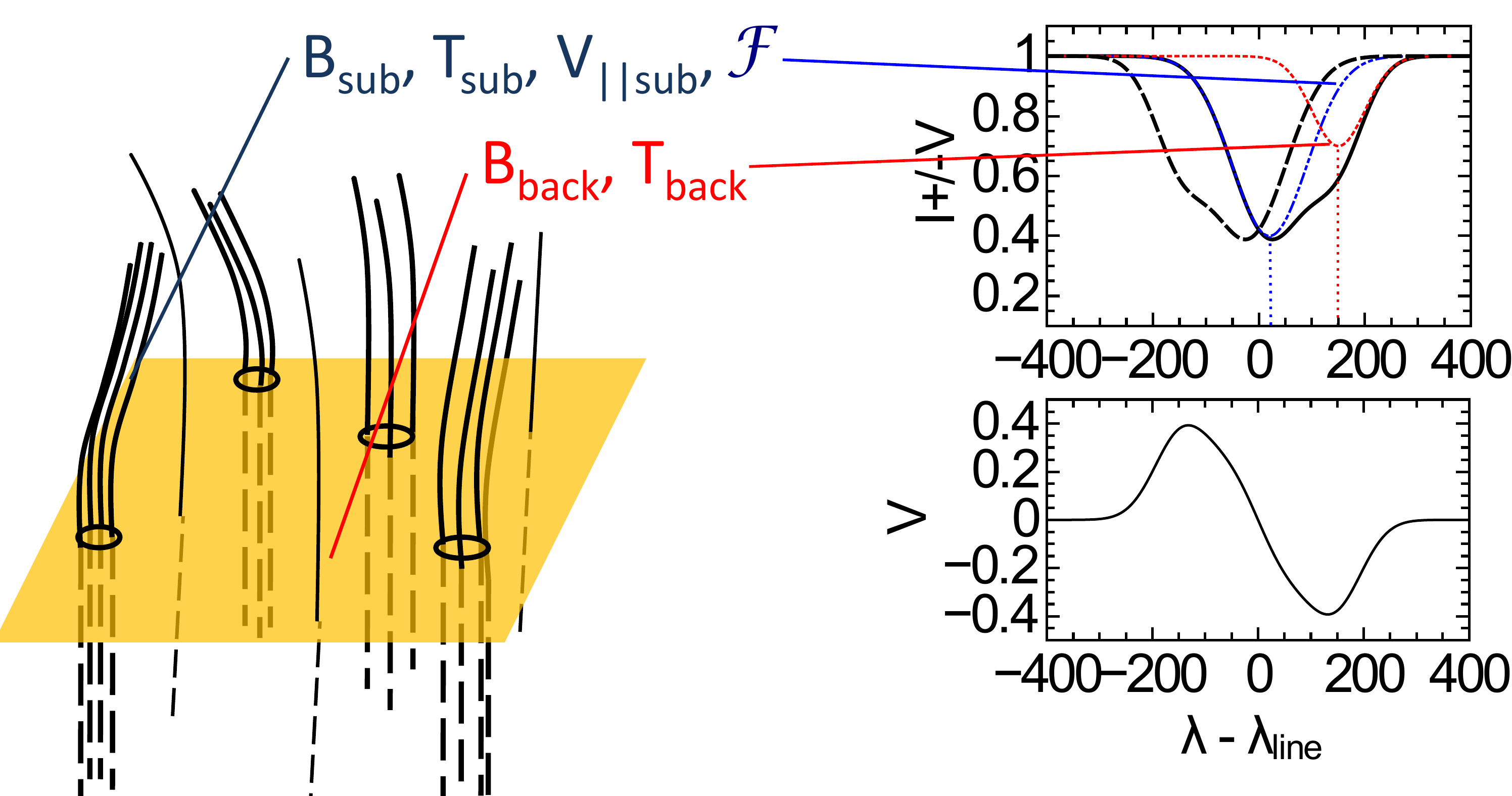}}
\caption{Left panel: two-component field with stronger unresolved magnetic flux tubes with field $B_\mathrm{sub}$ embedded into 
weaker ambient field $B_\mathrm{back}$, as shown in the left panel. Apart from different magnetic field strength, the 
two photospheric components may have different intensities and filling factors [$\mathcal{F}$], 
different widths (represented by their effective temperatures) and different line-of-sight (LOS) velocities. Right panels:
$I+V$ components corresponding to the background (blue dot-dashed line) and unresolved flux tubes (red dotted line) (C1 and C2, respectively), resulting $I+V$ and $I-V$ profiles (solid and dashed black lines, respectively). Lower panel shows corresponding Stokes-$V$ profile.}
\label{f-sketch}
\end{figure}

There is observational evidence that the photospheric magnetic field is very
inhomogeneous at small scales ({\it i.e.} $\approx 100$ km) (see, {\it e.g.} \opencite{sola93} for a review).
Early magnetographic observations showed that the magnetic field strengths evaluated using spectral
lines with similar characteristics but different magnetic sensitivity ({\it i.e.} different Lande factor [$g$]) can vary by up to a
factor of $2.5$ \cite{host72,sten73}. This effect was interpreted as an indicator of unresolved multi-component structure with
intense magnetic flux tubes embedded in a non-magnetic atmosphere or an atmosphere with weaker ambient field (Figure~\ref{f-sketch}). 
It is believed that 
these small-scale magnetic flux tubes may account for nearly $90\,\%$ of the photospheric magnetic flux outside sunspots \cite{frst72}.

Most existing instruments in the optical range can directly resolve features with sizes of $\approx 1$ Mm. Hence, here by 
``unresolved'' we mean spatial scales $\lesssim 1$~Mm.
The most direct measurements of unresolved field structure were carried out using speckle interferometry in the line 
Fe~{\sc I}~5250.2~$\ang$ \cite{keva92,kell92} and revealed magnetic elements with a field strength of a few kG and sizes of $100\,-\,200$~km, 
which, perhaps, can be considered as an upper limit on the diameters of small-scale magnetic flux tubes. Also, \citext{lin95}
observed full Stokes profiles in magneto-sensitive infrared Fe~{\sc I} lines 15648~$\ang$ and 15652~$\ang$ and found two types of small-scale 
magnetic elements: stronger elements with the field of $1.4$~kG and sizes $\approx 10^2\,-\,10^3$~km and weaker ones with the field strength
of $\approx 500$~G and sizes about $70$~km. The two types of magnetic elements were attributed to network and inter-network 
flux tubes. It was also concluded that the inter-network magnetic field elements have rather short lifetimes of about few hours.

There are a number of indirect measurements of unresolved magnetic field structure characteristics. 
Estimations of horizontal sizes of intense magnetic flux tubes vary significantly from tens of kilometers \cite{wieh78,lots89} to
hundreds of kilometers \cite{sanc98}. Comparison of the effective field values obtained using spectral lines with different magnetic sensitivity shows that the magnetic field in such flux tubes is about $1.0\,-\,3.0$~kG, although there are some indications that it could be substantially higher \cite{race05,lozi09}. The main reason behind such large discrepancies in estimations is that
even the two-component model, which is used to fit the observational data, has about ten free 
parameters, such as magnetic field strengths, field inclinations,
and surface brightness for both components, along with the filling factor and other parameters. Hence, diagnostics 
of the small-scale magnetic field require some realistic assumptions about the field structure to reduce the number of free parameters.
For instance, it might be safe to assume that the intense, unresolved flux tubes within a sampled area are almost identical \cite{ulre09}
and the key free parameters are the strength and vertical gradient of magnetic field in these flux tubes, and the filling factor.

The fine structure of the magnetic field in solar flares is understood even less than that in the quiet photosphere.
There is strong evidence of unresolved magnetic field in flares, but their diagnostics is a challenging task, because of the associated temperature and velocity inhomogeneities at small scales. Furthermore, it is not always possible to distinguish between
horizontal and vertical inhomogeneities at sub-telescopic scales.
\citext{lolo94} observed full Stokes profiles of several Fe~{\sc I} lines in order to study the structure of magnetic field in the 2B 
solar flare of 16~June~1989. It was found that the small-scale field strength was between $1.0$ and $1.5$~kG, and it substantially 
changed during the flare. It was also found that the filling factor decreased with time. 
More recently, the new generation of solar space observatories along with advanced ground-based instruments have provided more evidence for fine structure of the magnetic field in solar flares. Thus, full-Stokes-imaging spectropolarimetry of a C-class flare with the Interferometric Bidimensional Spectropolarimeter (IBIS) shows that Stokes profiles are highly irregular, indicating the presence of unresolved multi-component magnetic and velocity fields \cite{klei12}. The resolution of the instrument (up to 0.33~arcsec, or 240~km) provides the upper limit for the sizes of these unresolved magnetic elements. \citext{fise12} used the data from the Synoptic Optical Long-term Investigations of the Sun (SOLIS) vector-spectrograph in order to investigate the evolution of magnetic field in an X-class flare and also found that the Stokes profiles demonstrate complex, highly asymmetric structure that may be explained by a multi-component velocity field or by substantial perturbations of the spectral line profile due to heating. In addition, they show that a small patch of the photosphere, co-spatial with hard X-ray footpoints observed by Ramaty High Energy Solar Spectroscopic Imager (RHESSI), exhibits an unresolved fine structure. A quantitative estimation of a typical cross-section of small-scale magnetic elements can be made based on the analysis by \citext{anro12}. They carried out observations of the coronal rain using the Crisp Imaging Spectro-polarimeter (CRISP) at the Swedish Solar Telescope and concluded that the coronal rain consists of elements with a typical width of 
$\approx$310~km.
The structure and temporal variations of the 
small-scale magnetic field during flares may be related to the fast evolution of magnetic field in the corona and, therefore, 
the small-scale field structure in active regions and especially during solar flares, deserves more attention.

In the present work, we aim to study unresolved structure of magnetic field at the photosphere  during 
solar flares using two different approaches. The first approach is based on the analysis of the relationship between 
magnetic field strengths measured using different spectral lines and their Lande factors [$g$], which is similar to the method applied in magnetographic observations. The second 
approach is based on the analysis of fine structure of $I\pm V$ Stokes-profile splitting, which is possible only in observations 
with relatively high spectral resolution.


\section{Spectral Data and its Analysis} 
  \label{data}

\begin{figure}   
\centerline{\includegraphics[width=1.0\textwidth,clip=]{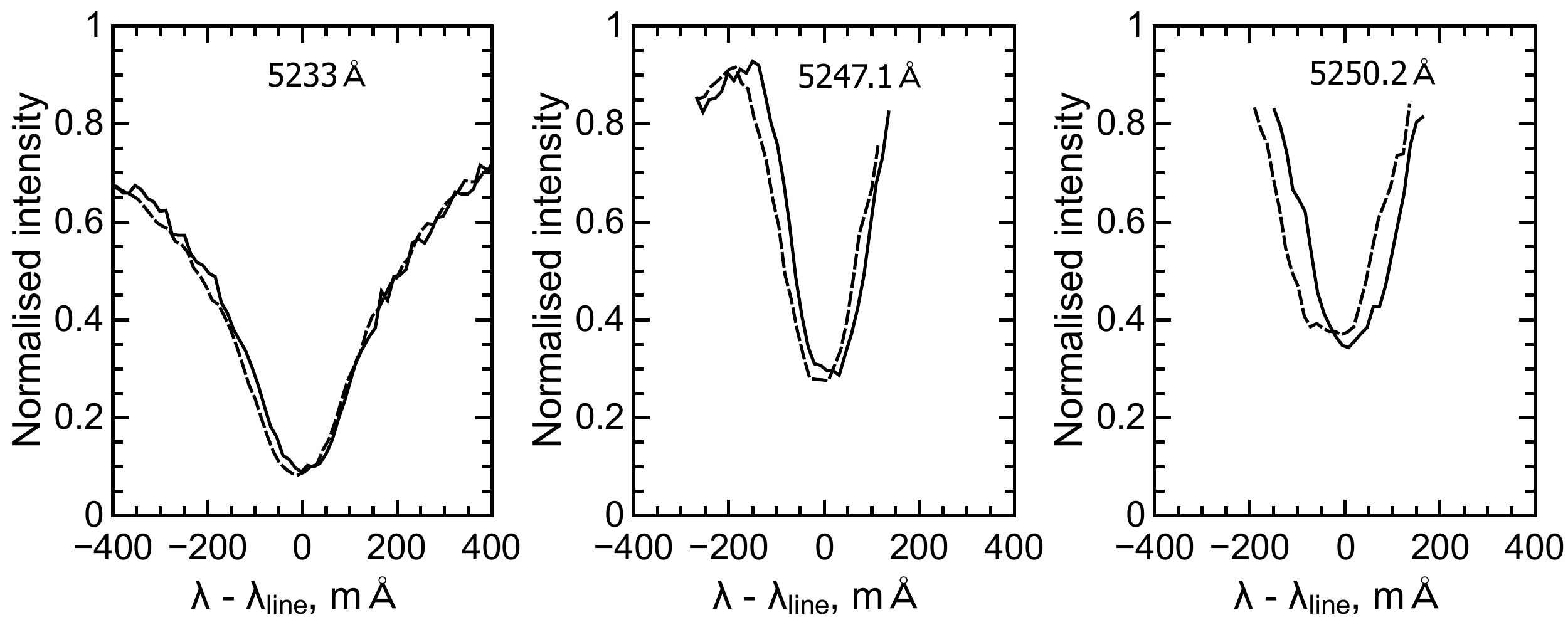}}
\caption{Typical $I+V$ (solid lines) and $I-V$ (dashed lines) Stokes profiles of Fe~{\sc I} 5233.0~$\ang$ (left panel), 5247.1~$\ang$ (middle panel), and 5250.2~$\ang$ (right panel) lines observed in a solar flare.}
\label{f-typical}
\end{figure}

\begin{figure}
\centerline{\includegraphics[width=0.55\textwidth,clip=]{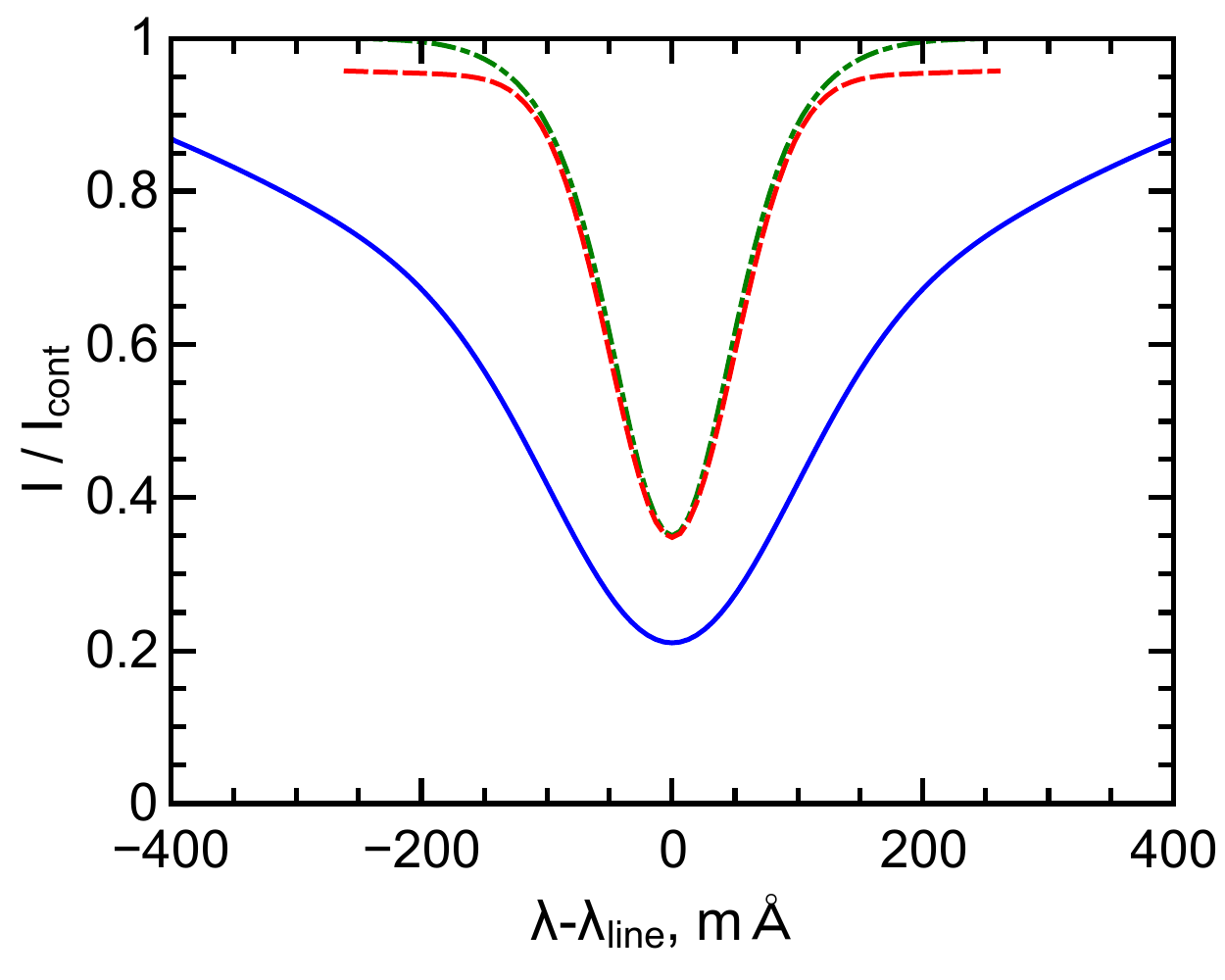}}
\caption{Smoothed and symmetrised profiles of Fe~{\sc I} 5233~$\ang$ (solid blue line), 5247.1~$\ang$ (dashed red line), and 5250.2~$\ang$ (dot-dashed green line) lines observed in plages away from sunspots.}
\label{f-lines}
\end{figure}
\begin{figure}    
\centerline{\includegraphics[width=0.55\textwidth,clip=]{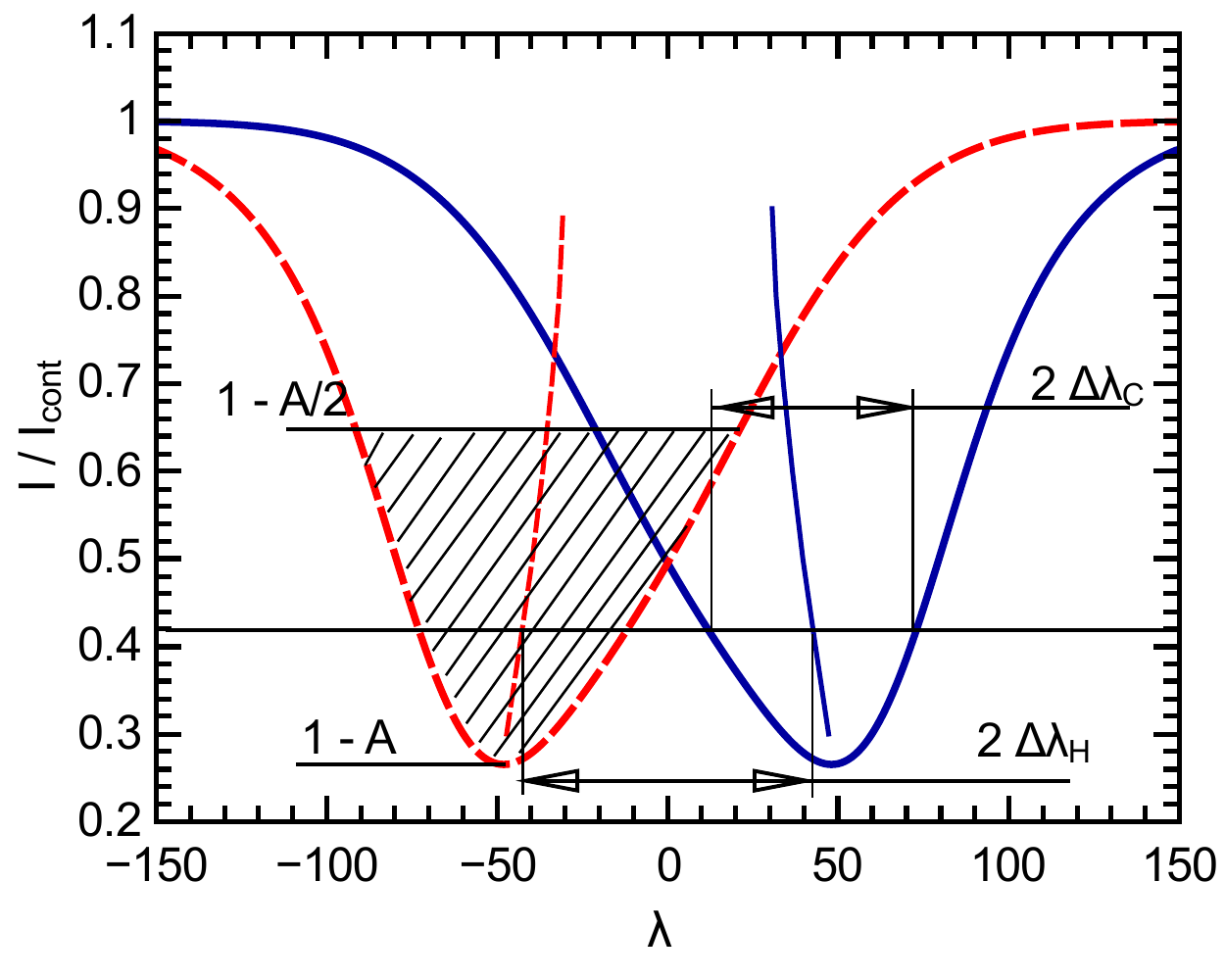}}
\caption{The scheme demontrates geometrical meaning of $\Delta \lambda_\mathrm{c}$, $\Delta \lambda_\mathrm{H}$ and $B_\mathrm{eff}$.
Blue solid and red dashed lines denote $I+V$ and $I-V$ components, respectively. Thin blue solid and thin red dashed lines denote bisectors corresponding to $I+V$ and $I-V$ Stokes profiles, respectively.
Hatched area within one of the components shows the part of a profile used to determine its centre-of-mass position, which, in turn, used
to deduce the value of $B_\mathrm{eff}$ (see Section ~\ref{observ-lr}).}
\label{f-scheme}
\end{figure}

\begin{figure}   
\centerline{\includegraphics[width=1.0\textwidth,clip=]{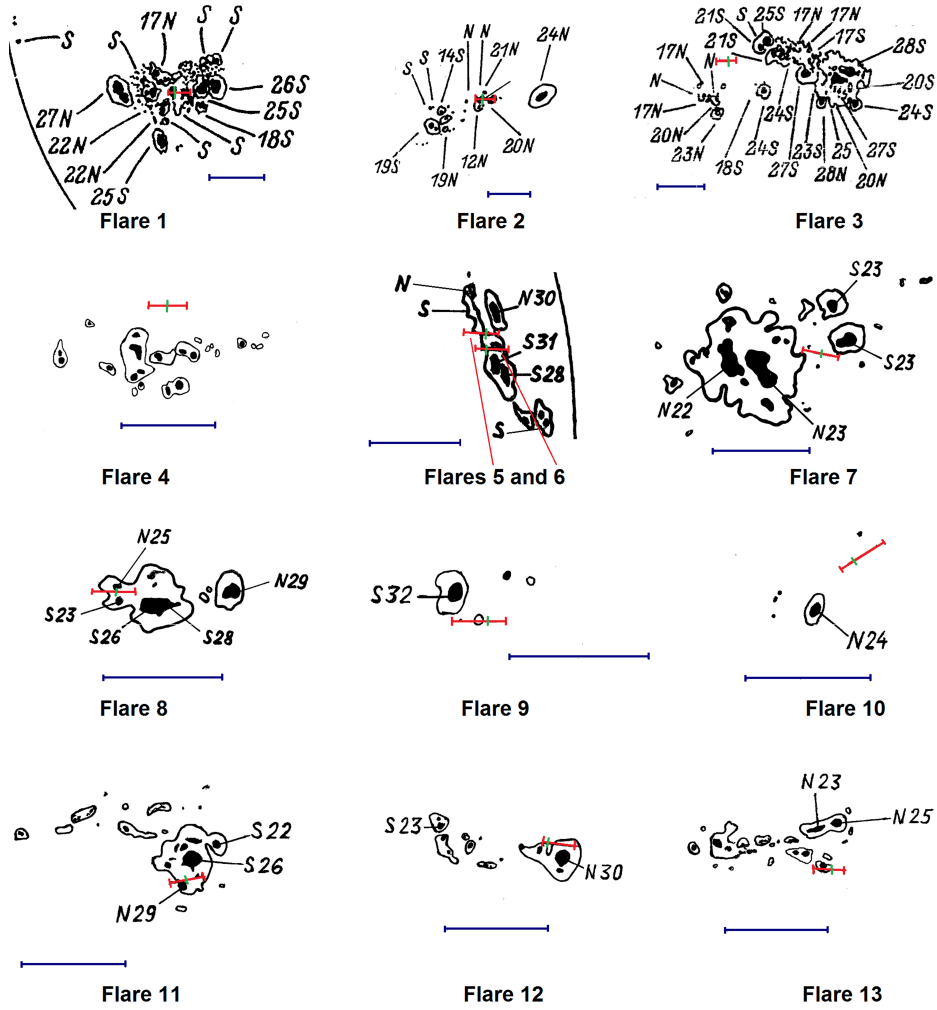}}
\caption{Active regions considered in the present study. Corresponding flare numbers are shown under each sketch. Blue scale corresponds to 2~arcmin on each sketch. Red lines show positions of the slit, while green dashes show the locations where the spectra have been taken.}
\label{f-sketches}
\end{figure}

Observations of 13 solar flares of different classes are analysed; their details are given in Table~\ref{tabobj}. For comparison, we 
also present observations of a plage and a sunspot. The observational data were obtained on the 
echelle spectrograph of the horizontal solar telescope of the Kyiv National 
University \cite{kure80}. This spectrograph can simultaneously record a spectrum in the range from 3800~$\ang$ to 6600~$\ang$ 
with spectral resolution of about 30 m$\ang$ in the green part of the spectrum. The spatial resolution is about $2-3$~arcsec, depending on atmospheric conditions, which means that the observed spectra correspond to areas of about $3 - 4$~Mm$^2$.
The exposure time was $15-20$~seconds.

All of the spectra have been observed during the main (or ``thermal'') phase of flares (see Table 1) at the locations with brightest H$\alpha$ emission. Specific slit locations are shown in Figure~\ref{f-sketches}.

\begin{table}[!ht]
\caption{Solar flares considered in the present study. Columns from the left to the right show: date, time the spectrum was taken, onset time in X-rays, time of maximum intensity in H$\alpha$, flare location on the disc, flare class. The onset time, maximum time, location and class for flares 4~--~13 are taken from NOAA solar event reports (\href{http://www.swpc.noaa.gov}{www.swpc.noaa.gov}). The spectra correspond to the locations shown in Figure~\ref{f-sketches}.}
\label{tabobj}
\begin{tabular}{lllllll}
\hline      
& Date & Time UT & Start UT & Max UT & Location & Class\\
\hline
1 & 25 Jul. 1981 & 12:58	& ??:?? & ??:?? & N11E36 & 2N \\
2 & 15 Jun. 1989 & 11:29	& ??:?? & ??:?? &  N20E10 & 1B \\
3 & 16 Jun. 1989 & 09:30	& ??:?? & ??:?? &  S17E04 & 2B \\
4 & 14 Jul. 2000 & 13:53	& 13:44 & 13:50 &  N20W08 & M3.7/1N  \\
5 & 02 Apr. 2001 & 10:07 & 10:04 & 10:07 &  N17W60 & X1.4/1B \\
6 & 02 Apr. 2001 & 12:04 & 10:58 & ??:?? &  N17W60 & X1.1/3N \\
7 & 28 Oct. 2003 & 11:13 & 09:51 & 12:05 &  S16E08 & X17.2/4B \\
8 & 05 Nov. 2004 & 11:37 & 11:23 & 11:29 &  N08E15 & M4/1F \\
9 & 03 Aug. 2005 & 14:09 & 13:48 & 14:07 &  S14E36 & C9.3/1N \\
10 & 07 May 2012 & 14:28 & 14:03 & 14:25 &  S19W46 & M1.9/1N \\
11 & 10 May 2012 & 13:58 & 13:10 & 13:47 &  N07E09 & C5/SF \\
12 & 13 Jun. 2012 & 13:25 & 11:29 & 13:41 &  S16E18 & M1.2/1N \\
13 & 02 Jul. 2012 & 11:00 & 10:43 & 10:52 &  S17E08 & M5.6/2B \\
\hline
\end{tabular}
\end{table}

\begin{table}[!ht]
\caption{Spectral lines used in observations}
\label{tablines}
\begin{tabular}{|l||c|c|c|c|c|c|c|c|}
\hline      
Element & Fe {\sc I} & Fe {\sc I} & Fe {\sc I} & Fe {\sc I} & Fe {\sc I} & Fe {\sc I} & Cr {\sc I} & Fe {\sc I} \\
&&&&&&&&\\
$\lambda$, [$\ang$] & 5123.7 & 5434.5 & 5576.1 & 5233.0 & 5250.6 & 5247.1 & 5247.6 & 5250.2 \\
&&&&&&&&\\
g factor & -0.01 & -0.01 & -0.01 & 1.26 & 1.50 & 2.00 & 2.50 & 3.00 \\
&&&&&&&&\\
\hline
\end{tabular}
\end{table}

We analyse $I\pm V$ Stokes profiles of spectral lines of neutral iron and neutral chromium with different magnetic sensitivity 
(see Table \ref{tablines}). Several ``non-magnetic lines'' ({\it i.e.}, lines with very low Lande [$g$] factor) have also been analysed in order to distinguish between magnetic and non-magnetic effects. In addition, observations of the telluric line H$_2$O
$\lambda=5919.6$~$\ang$ are taken into account to eliminate atmospheric effects and evaluate the characteristic error 
in our spectral measurements.

The choice of spectral lines is typical for this type of observations; they have been extensively used to measure magnetic fields 
both in quiet regions and in flares \cite{sten73,sole87,lolo94,khco07}. We particularly focus on the Fe~{\sc I}~5233~$\ang$, which has lower temperature
and velocity sensitivity due to its width, compared to other ``classical'' lines: Fe~{\sc I}~5247.1~$\ang$ and 5250.2~$\ang$ \cite{frst72}. 

Our analysis of the observed Stokes profiles and synthetic profiles is based upon the assumption that the field can be described 
using two-component configuration, and both components contribute to all spectral lines used in observations. This assumption requires all of the considered lines to have nearly equal formation depths.
The formation depths of 5247.1~$\ang$ and 5250.2~$\ang$ are $\approx 320-330$~km, which is lower than the formation depth of the 5233~$\ang$ line -- $\approx 400$~km
\cite{kostykbook}. However, all of these lines have extended formation heights spanning up to $200-300$~km, which is larger than the difference in average formation depths (see \citext{khco07} and discussion therein). Hence, it is acceptable to assume that these lines sample approximately the same range of heights in the photosphere. 

Typical profiles of three Fe~{\sc I} lines observed in flares are shown in Figure~\ref{f-typical}, while Figure~\ref{f-lines} shows their ``averaged'' smoothed profiles observed in plages. The latter profiles will be used for synthetic profile calculations in Section~\ref{synt}).
The main difference between the studied lines is their half-widths: in active regions the 5233~$\ang$ line has $\Delta \lambda_{1/2} \approx 180$ m$\ang$ compared to the 5247.1~$\ang$ and 5250.2~$\ang$ lines with $\Delta \lambda_{1/2} \approx 70$ m$\ang$. Hence, taking into account the difference in their magnetic sensitivity ($\Delta \lambda_\mathrm{H} /B$ = 16.1 m$\ang$/kG for the 5233~$\ang$, 25.7 m$\ang \, kG^{-1}$ for the 5247.1~$\ang$, and 38.6 m$\ang \, kG^{-1}$ for the 5250.2~$\ang$), even magnetic field of $1-2$~kG would lead to full Zeeman splitting of the 5247.1~$\ang$ and 5250.2~$\ang$ pair of lines, while in the case of the 5233~$\ang$ line the Zeeman splitting remains smaller than the line half-widths up to $\approx 10$~kG.


\section{I$\pm$V Stokes Profiles Observed in Solar Flares}
	\label{observ}

\subsection{Magnetic Field Deduced From Spectral Lines with Different Magnetic Sensitivities}
	\label{observ-lr}

\begin{figure}[!ht]    
\centerline{\includegraphics[width=0.55\textwidth,clip=]{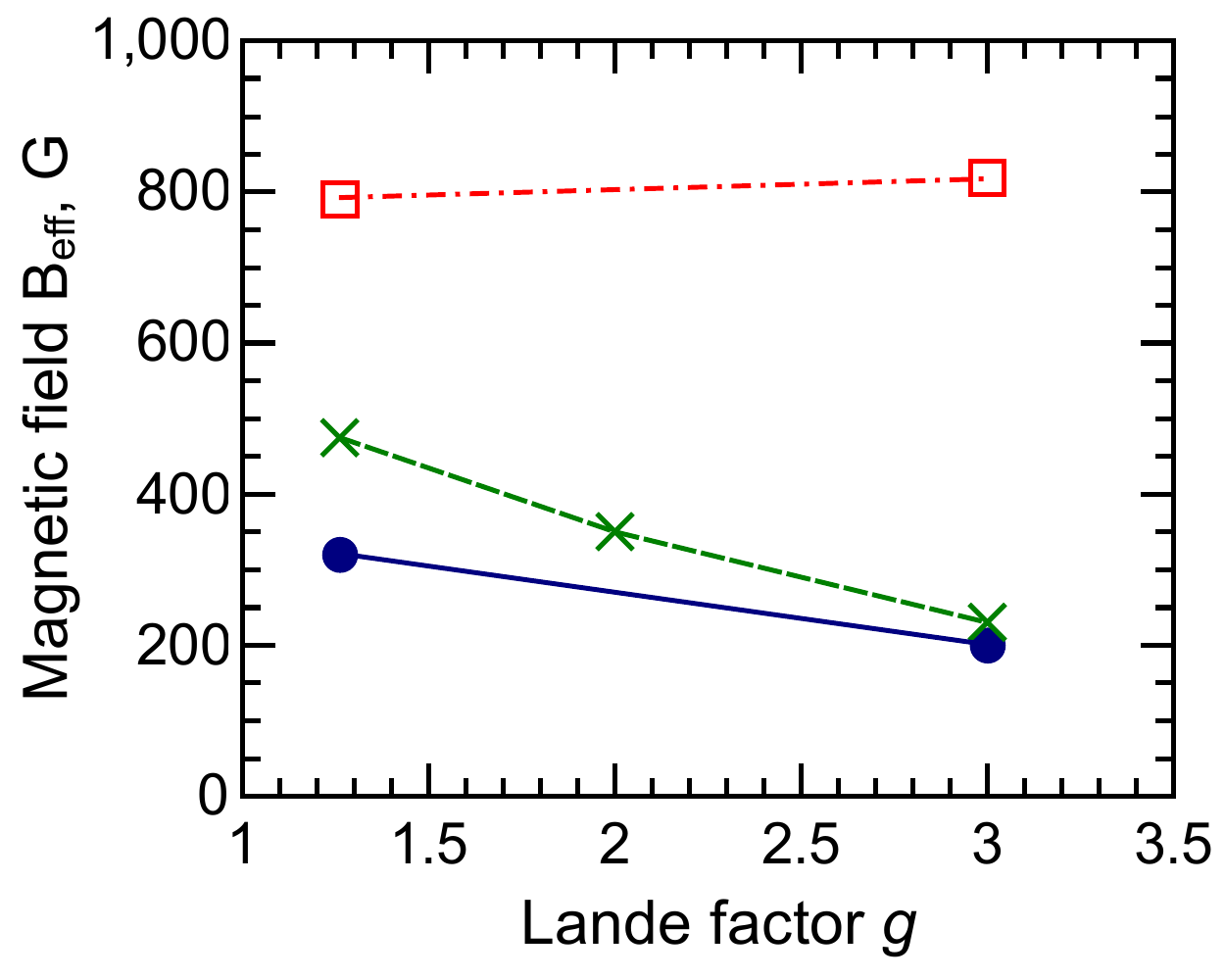}}
\caption{Typical relations between the effective magnetic field strength and Lande factor $g$ in the quiet photosphere (blue solid line with solid circles), in a plage (green dashed line with crosses), and a sunspot umbra (red dot-dashed line with open squares).}
\label{f-bvg-quiet}
\end{figure}
\begin{figure}[!h]    
\centerline{\includegraphics[width=0.55\textwidth,clip=]{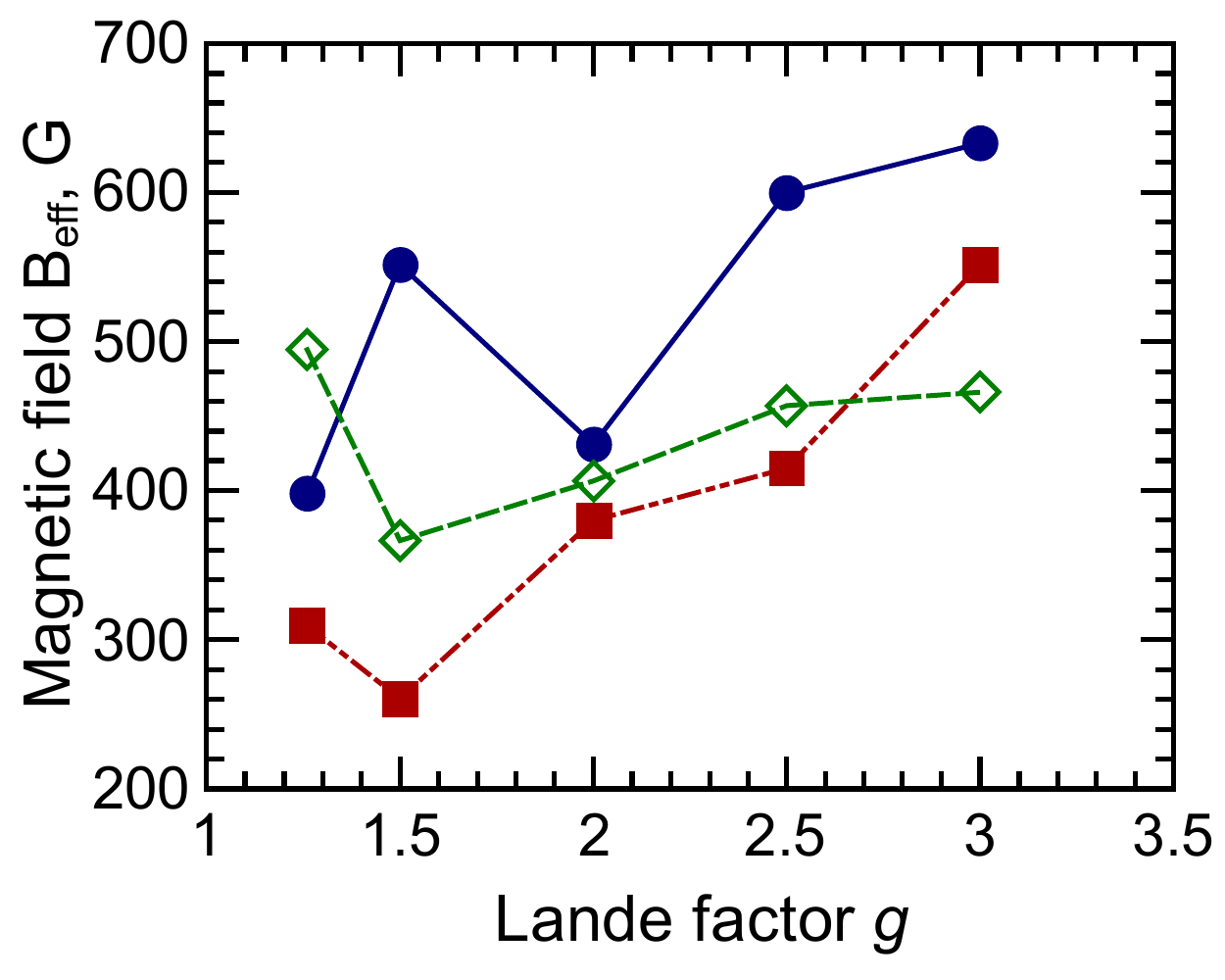}}
\caption{Relations between the effective magnetic field strength and Lande factor $g$ in flaring photospheres. Solid blue, dot-dashed red and dashed green lines correspond to flares 5, 6, and 11, respectively.}
\label{f-bvg-flare}
\end{figure}

Comparison of effective magnetic field strengths obtained using spectral lines with different magnetic sensitivity is the 
main method used to analyse spatial magnetic field inhomogeneity. The magnitude of line splitting due to Zeeman effect 
is related to the magnetic field as   
\begin{equation}\label{eq-zeeman}
\Delta \lambda_\mathrm{H} = \mathcal{K} g \lambda^2 B
\end{equation} 
where $\Delta \lambda_\mathrm{H}$ is splitting between $\pi$- and $\sigma$-components and $\mathcal{K}$ is a constant, which is equal 
to $\mathcal{K}=4.67\times 10^{-13}$ if [$\lambda$] and [$\Delta \lambda_\mathrm{H}$] are measured in $\ang$ and the magnetic 
field [$B$] is in Gauss. 
In this section we compare observed effective magnetic field values [$B_\mathrm{eff}$] deduced from different spectral lines. Here $B_\mathrm{eff}$ is defined using Equation~(\ref{eq-zeeman}) with $\Delta \lambda_\mathrm{H}$ corresponding to half of the distance between the centres-of-mass of $I+V$ and $I-V$ profiles (see Figure~\ref{f-scheme}). Hence, $B_\mathrm{eff}$ represents some volume-averaged value of the line-of-sight magnetic field component. 

Figure~\ref{f-bvg-quiet} shows magnetic field  
$B_\mathrm{eff}$ as a function of Lande [$g$] factor for a typical sunspot and a typical plage, while Figure~\ref{f-bvg-flare} shows 
$B_\mathrm{eff}(g)$ for three solar flares. 

In sunspots, the field values given by different lines are quite close, indicating rather solid homogeneous magnetic field structure. Outside sunspots the observed value of magnetic field is normally lower for lines with higher Lande factor. Thus, in 
quiet photosphere and in plages, the field strength
measured with the $5233$~$\ang$ line is higher by factor of $1.5-2.0$ than the field strength measured using the $5250.2$~$\ang$ line. 
In contrast, observations of magnetic field in solar flares reveal the opposite picture: the effective field is higher 
for higher Lande factors. It can be seen, that the field observed with lines with $g <2$ is in the range of $300-500$~G, 
while more sensitive lines with $g > 2.5$ yield values in the range $500-700$~G.

Observations of the quiet photosphere are in good agreement with previous data. These can be explained by the 
saturation effect, resulting from the presence of unresolved field of the same polarity \cite{frst72,ulre09}. 
The essence of the saturation effect is that 
the contribution of spectral component with high Zeeman splitting increases the effective splitting of $I\pm V$ components when 
$\Delta \lambda_\mathrm{H\;  sub}$ corresponding to the strong field is smaller than the width of spectral line $\Delta \lambda_{1/2}$, but when 
$\Delta \lambda_\mathrm{H\;  sub} > \Delta \lambda_{1/2}$ the effective field strength decreases with the increase of $\Delta \lambda_\mathrm{H\;  sub}$. 
The saturation effect will be discussed in more details in Section~\ref{synt-ratio}.
The increase of $B_\mathrm{eff}$ with $g$ observed in solar flares cannot be explained by the normal 
saturation effect. However, there are three alternative
scenarios, that can explain the observed phenomenon based on the two-component field model: firstly, 
the field strenth in unresolved magnetic elements (C2 component) can be lower than the background field. Secondly, the unresolved magnetic elements may have polarity opposite to that of the background field. Finally, stronger unresolved 
magnetic elements may produce emission. In all these cases, the contribution of C2 component would lead to lower effective field value $B_{\mathrm eff}$. That contribution will be bigger for low $g$ and smaller for high $g$, hence, resulting in $B_\mathrm{eff}$ increasing with
$g$.
In general, all 
these scenarios are viable: indeed, solar 
flares normally occur in active regions with strongly mixed magnetic field polarities and, at the same time, 
many spectral lines in solar flares
demonstrate noticeable emission peaks within their absorption profiles. In 
order to distinguish between these two
scenarios, we also analyse the detailed structure of spectral line splitting (Section~\ref{observ-bs}). Then, 
we calculate synthetic profiles of $5233.0$~$\ang$,
$5247.1$~$\ang$, and $5250.2$~$\ang$ lines for different two-component field configurations in order to determine which configuration provides the 
best fit for the observational data (see Section ~\ref{synt}). 

\subsection{Fine Structure of I$\pm$V Bisector Splitting}
	\label{observ-bs}

\begin{figure}[!hb]    
\centerline{\includegraphics[width=0.55\textwidth,clip=]{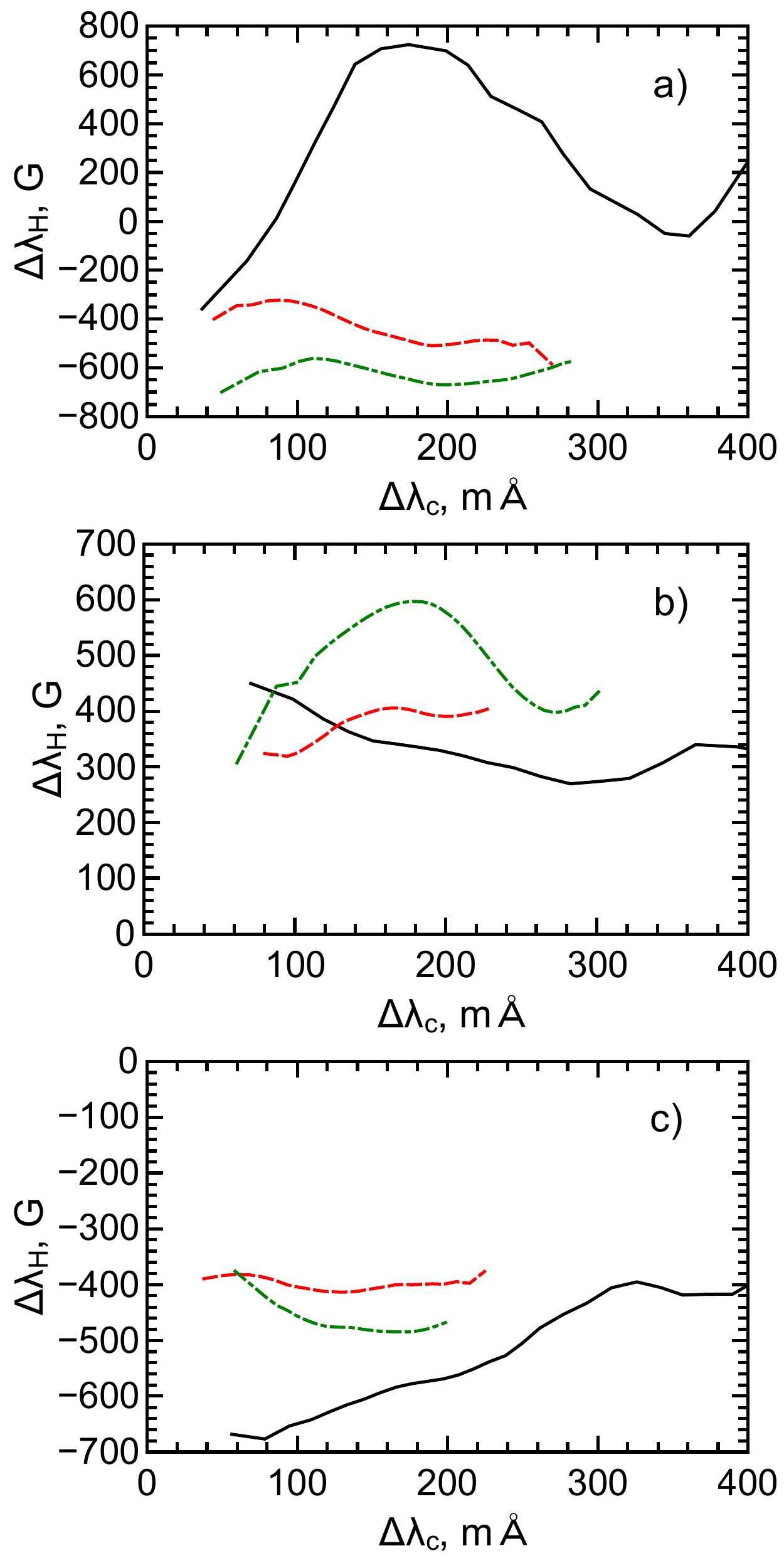}}
\caption{Bisector splitting functions for Fe~{\sc I} 5233~$\ang$ (solid black lines), 5247.1~$\ang$ (red dashed lines) and 5250.2~$\ang$ 
(green dot-dashed lines) for flares 5 (panel a), 6 (panel b), and 11 (panel c).}
\label{f-bsall}
\end{figure}
\begin{figure}   
\centerline{\includegraphics[width=0.95\textwidth,clip=]{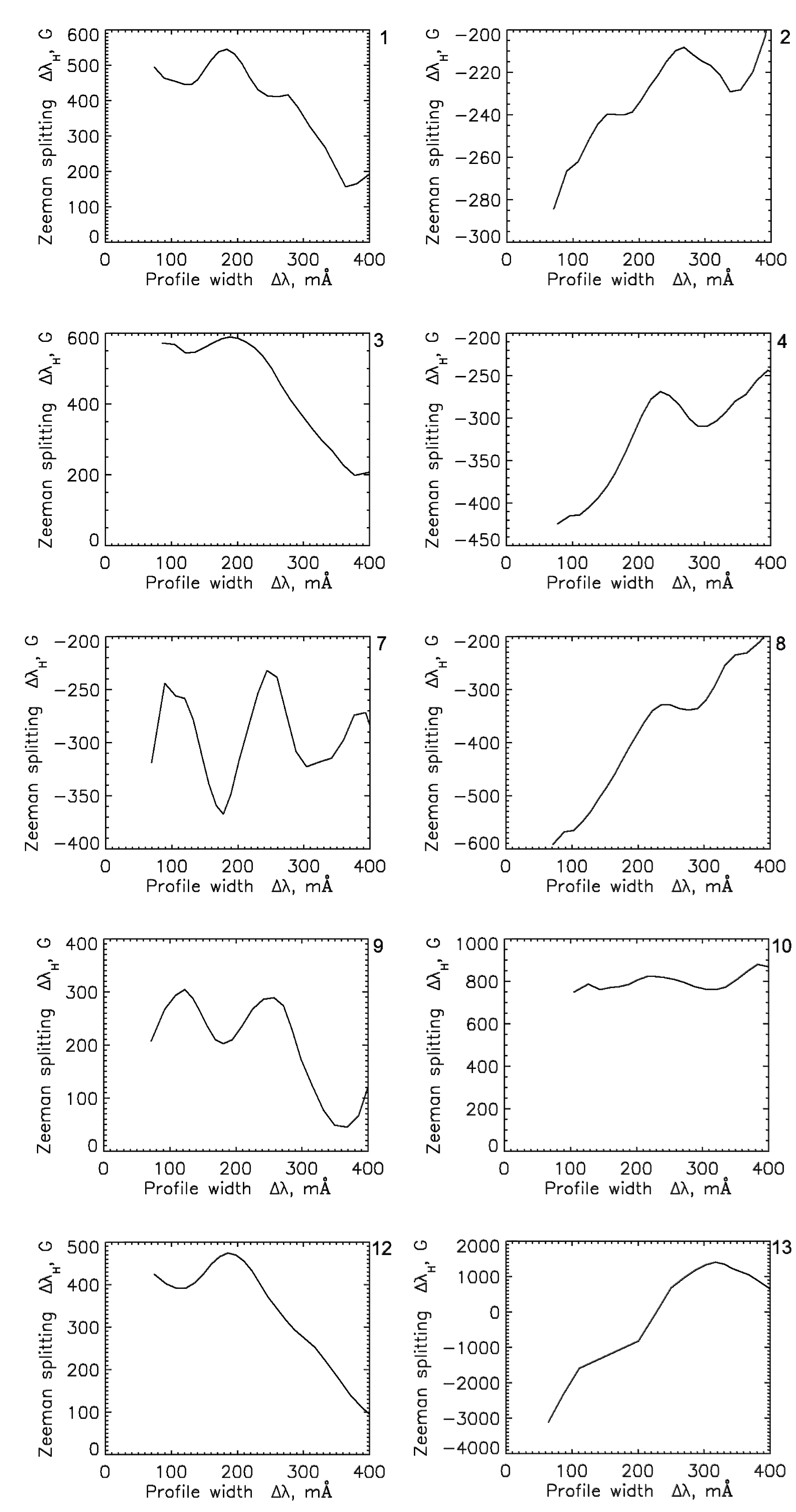}}
\caption{Bisector splitting functions of 5233~$\ang$ line for the flares not shown in Figure~\ref{f-bsall}.}
\label{f-bs5233}
\end{figure}
\begin{figure}[!h]    
\centerline{\includegraphics[width=0.55\textwidth,clip=]{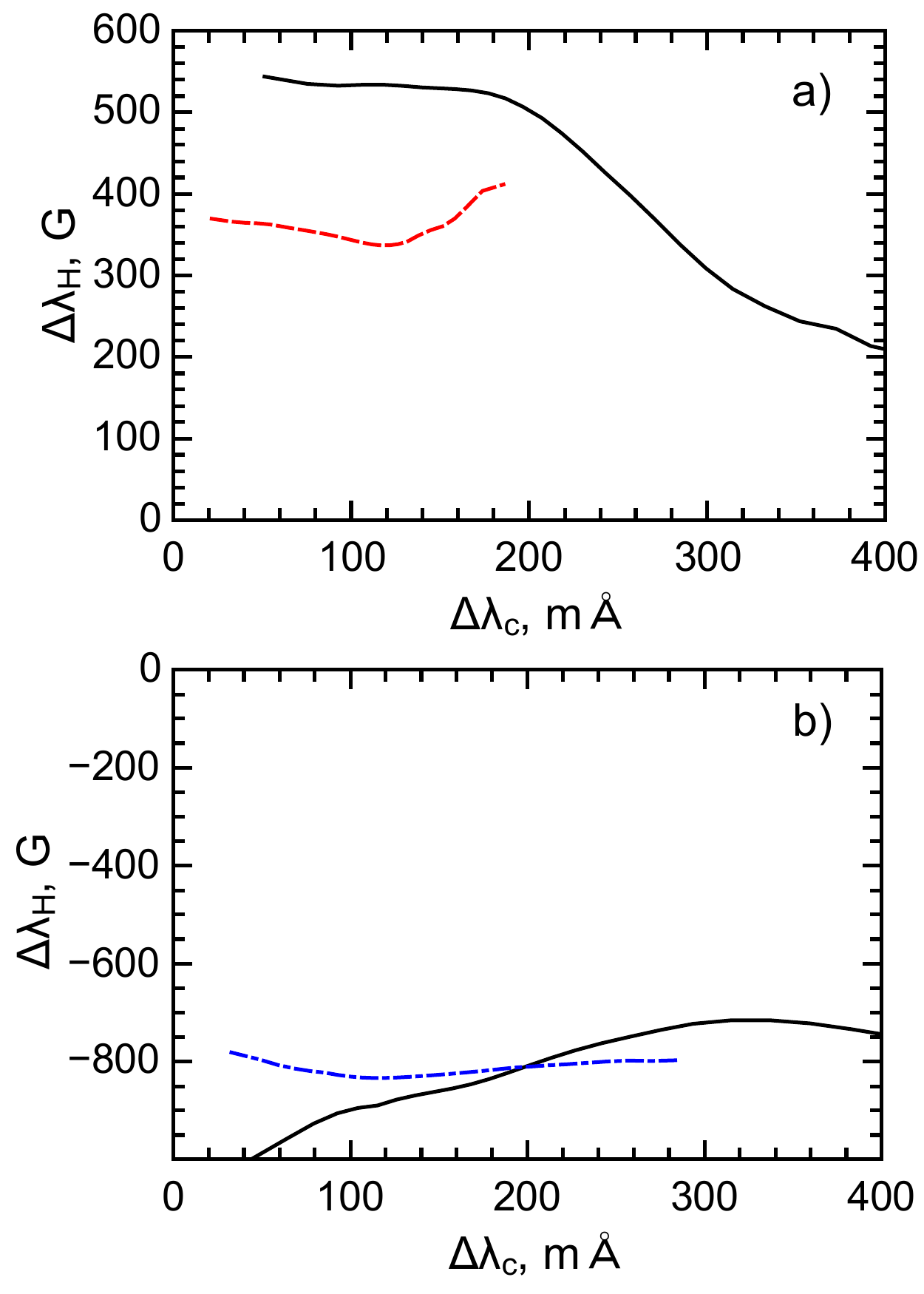}}
\caption{Panel a: Bisector splitting functions for a typical plage, solid black and dashed red lines correspond to Fe~{\sc I} 5233~$\ang$ 
and 5250.2~$\ang$ lines. Panel b: Bisector splitting functions for a typical sunspot penumbra, solid black and dot-dashed blue lines correspond to Fe~{\sc I} 5233~$\ang$ and 5250.6~$\ang$ lines, respectively.}
\label{f-bsother}
\end{figure}

Generally, different components of the magnetic field within the same spatially unresolved area of the photosphere may penetrate plasmas with different 
flow velocities, temperatures, turbulent velocities {\it etc} (see Figure~\ref{f-sketch}). Hence, spectral 
components corresponding to the unresolved flux tubes and ambient magnetic field, apart from different  
Zeeman splitting $\Delta \lambda_\mathrm{H}$ and surface intensities, may have different Doppler widths $\Delta \lambda_{1/2}$ and different Doppler shifts $\Delta \lambda_V$. The combination of all these factors may result 
in different Zeeman splitting of the cores and wings of spectral lines. This effect can be described by the value of $I\pm V$ 
bisector splitting $\Delta \lambda_\mathrm{H}$ 
measured as a function of the distance from the line centre $\Delta \lambda_\mathrm{c}$. The latter is the width of the profile measured at a given intensity level 
(Figure~\ref{f-scheme}, see also Section  4.3 in \opencite{ulre09}).

Bisector splitting functions $\Delta \lambda_\mathrm{H} (\Delta \lambda_\mathrm{c})$ of Fe~{\sc I} 5233~$\ang$, 5247~$\ang$, and 5250~$\ang$ lines for three flares are shown in 
Figure~\ref{f-bsall}. Additionally, Figure~\ref{f-bs5233} shows the $\Delta \lambda_\mathrm{H} (\Delta \lambda_\mathrm{c})$ function for 5233~$\ang$ line for ten more flares. For comparison,
Figure~\ref{f-bsother} shows typical bisector splitting functions for a sunspot and a plage.

It can be seen that bisector splitting functions corresponding to a sunspot do not show substantial variations, again indicating rather uniform magnetic field. At the same time, in a plage the bisector splitting functions show a slow decrease with $\Delta \lambda_\mathrm{c}$, similar to Zeeman splitting observed in the quiet photosphere. This cannot be explained by normal vertical field gradient, as it should yield an opposite picture: the line 
cores are formed in colder regions closer to the temperature minimum where the field is normally weaker, while wings are 
formed deeper regions, where the field is normally stronger. An alternative explanation, with the magnetic field being mildly inhomogeneous in the
horizontal direction due to multi-thread field structure, seems to be a more realistic explanation.

In flares, the structure of bisector splitting is more complicated. In most flares the general trend is similar to the quiet photosphere: the value 
of $\Delta \lambda_\mathrm{H}$ almost linearly decreases from cores to wings. However, in some flares (5, 7, 9, 11) there are substantial 
deviations from the trend in the form of narrow minima and maxima; these deviations are greater than the typical error of $\Delta \lambda_\mathrm{H}$ measurements, 
$\approx 150$ G, see Section~\ref{synt-error}). At the same time, in one of the considered flares [10] the trend is nearly horizontal.

The general trend of 
$\Delta \lambda_\mathrm{H}(\Delta \lambda_\mathrm{c})$ in flares can be easily explained by the same two factors: field convergence and mild horizontal inhomogeneity. 
However, in order to explain localised deviations from the trend it is necessary to assume that there are one or more components of spectral lines
with the widths considerably smaller that the width of the main (or background) absorption component (Section ~\ref{synt-bs}). Indeed, such features are often observed in flares: 
thus, \citexts{loze99}{loze00} reported observations of very narrow emission components appearing in cores of some Fe~{\sc I} lines. In principle, if the
magnetic splitting of these components is different from the magnetic splitting of the main absorption component, the resulting $\Delta \lambda_\mathrm{H}(\Delta \lambda_\mathrm{c})$
may show complex profiles similar to those in Figure \ref{f-bsall}; this will be considered in the next section.


\section{Synthetic I$\pm$V Stokes Profiles Based on Two-component Magnetic Field Models}\label{synt}

In order to interpret our observational data, synthetic $I\pm V$ Stokes profiles are calculated for three spectral lines -- Fe~{\sc I} 5233.0~$\ang$, 5247.1~$\ang$, and 5250.2~$\ang$  -- based on the two-component magnetic field model. 

\subsection{Calculation of Synthetic Profiles}\label{synt-method}

The resulting $I\pm V$ spectra are assumed to be linear superpositions of background spectra $(I\pm V)_\mathrm{back}$ (or C1 component) and spectra corresponding to the small-scale magnetic elements $(I\pm V)_\mathrm{sub}$ (or C2 component):

\begin{equation}\label{eq-combi}
(i\pm v) (\lambda) = (1-\mathcal{F}) [i\pm v]_\mathrm{back} (\lambda) + \mathcal{F} [i\pm v]_\mathrm{sub} (\lambda),
\end{equation}
Here $i$ and $v$ denote Stokes parameters normalised by the continuum intensity $I_{cont}$: 
$i(\lambda) = I(\lambda)/I_{cont}$, $v(\lambda) = V(\lambda)/I_{cont}$. The filling factor $\mathcal{F}$ is used to account for the differences in the
surface brightness and surface area.

Plages in active regions have thermodynamic conditions most similar to those in flaring photosphere and, 
therefore, we use smoothed and symmetrised line profiles observed in plages $i_\mathrm{plage}(\lambda)$ (see Figure~\ref{f-lines})
as the background spectral component (Component 1, see Figure~\ref{f-sketch}). Hence, the background 
component is defined as 

\begin{equation}\label{eq-comp1}
\begin{array}{l}
(i+v)_\mathrm{back} (\lambda) = i_\mathrm{plage}(\lambda -\Delta \lambda_\mathrm{H\;back})\\
(i-v)_\mathrm{back} (\lambda) = i_\mathrm{plage}(\lambda +\Delta \lambda_\mathrm{H\;back}),
\end{array}
\end{equation}
where $\Delta\lambda_\mathrm{H\;back}$ is Zeeman shift of Component 1 $\sigma$-components, as defined by Equation~\ref{eq-zeeman}

Next we consider several possible scenarios with different spectral manifestations corresponding to the
strong field component (Component 2). Thus, the component (of the considered spectral line) corresponding to the
Component 2 may have either absorption or emission profile. In both cases we assume that $(i\pm v)_\mathrm{sub}$ profiles have Gaussian 
shapes, and, hence the Component 2 profiles are defined as follows:
\begin{equation}\label{eq-comp2}
\begin{array}{l}
(i+v)_\mathrm{sub}(\lambda) = 1 - a \exp(-(\lambda -\lambda_\mathrm{line} - \Delta\lambda_\mathrm{H\; sub}  - \Delta \lambda_\mathrm{LOS})^2/(\Delta \lambda_{1/2})^2)  \\
(i-v)_\mathrm{sub}(\lambda) = 1 - a \exp(-(\lambda -\lambda_\mathrm{line} + \Delta\lambda_\mathrm{H\; sub}  - \Delta \lambda_\mathrm{LOS})^2/(\Delta \lambda_{1/2})^2), 
\end{array}
\end{equation}
Here $\Delta\lambda_\mathrm{H\; sub}$ is the Zeeman shift of the Component 2 $\sigma$-components, 
$\Delta \lambda_\mathrm{LOS}$ is Doppler shift of C2 due to non-zero line-of-sight (LOS) velocity in the regions where C2 is formed, and $\Delta \lambda_{1/2}$ is the half-width of the Component 2 profiles, which is equal to the half-width of the corresponding background profile, unless otherwise stated. We also measure the half-width in terms of corresponding equivalent temperature $T_\mathrm{sub}$: 
\begin{equation}\label{eq-dopwid}
\Delta \lambda_{1/2} = \frac{\lambda_\mathrm{line}}{c} \sqrt{ \frac{2 k_B T_\mathrm{ sub}}{m_\mathrm{Fe}} }, 
\end{equation}
although it should be noted that it accounts not only for the thermal broadening, but also for the turbulence and other factors.
The parameter $a$ in Equation~\ref{eq-comp1} is equal to 1 in the case of C2 in absorption and is equal to -1 when C2 is in emission.

The assumption about Gaussian shapes of the Component 2 profiles is definitely viable in case of emission, as it is normally formed
in an optically thin layer above the temperature minimum. In the case of absorption, the line profiles are most likely optically thick. However, this should
not result in substantial errors, as we are interested mostly in the contribution of cores, which have shapes very close to Gaussian.

The obtained synthetic profiles are analysed in the same way as the observed ones: we derive the effective magnetic field 
$B_\mathrm{eff}(g)$ and bisector splitting functions $\Delta \lambda_\mathrm{H}(\Delta \lambda_\mathrm{c})$.

\begin{figure}[!h]    
\centerline{\includegraphics[width=0.55\textwidth,clip=]{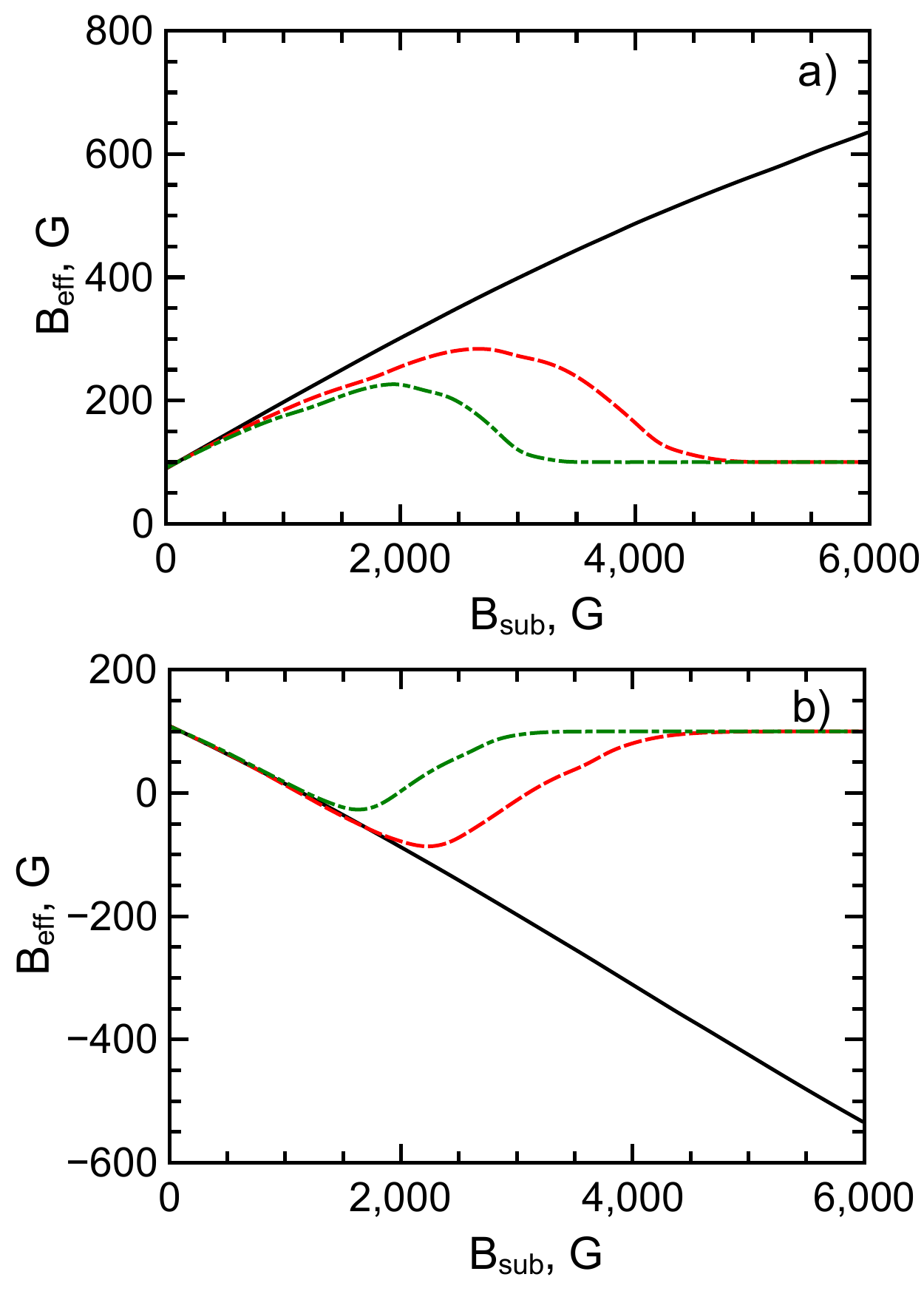}}
\caption{Effective field $B_\mathrm{eff}$ versus $B_\mathrm{sub}$ for Fe~{\sc I} 5233~$\ang$ (solid black line), 5247.1~$\ang$ (dashed red) and 5250.2~$\ang$ (dot-dashed green). 
Panel a is for the case of C2 in absorption, panel b is for C2 in emission. In both cases C2 spectral component has the same width as
the main component. Here $B_\mathrm{back}= 100$~G and the filling factor is $\mathcal{F}=0.1$.}
\label{f-saturation}
\end{figure}

\subsection{Effective Magnetic Field Values Based on the Synthetic Profiles}\label{synt-ratio}

\begin{figure}[!ht]    
\centerline{\includegraphics[width=1.0\textwidth,clip=]{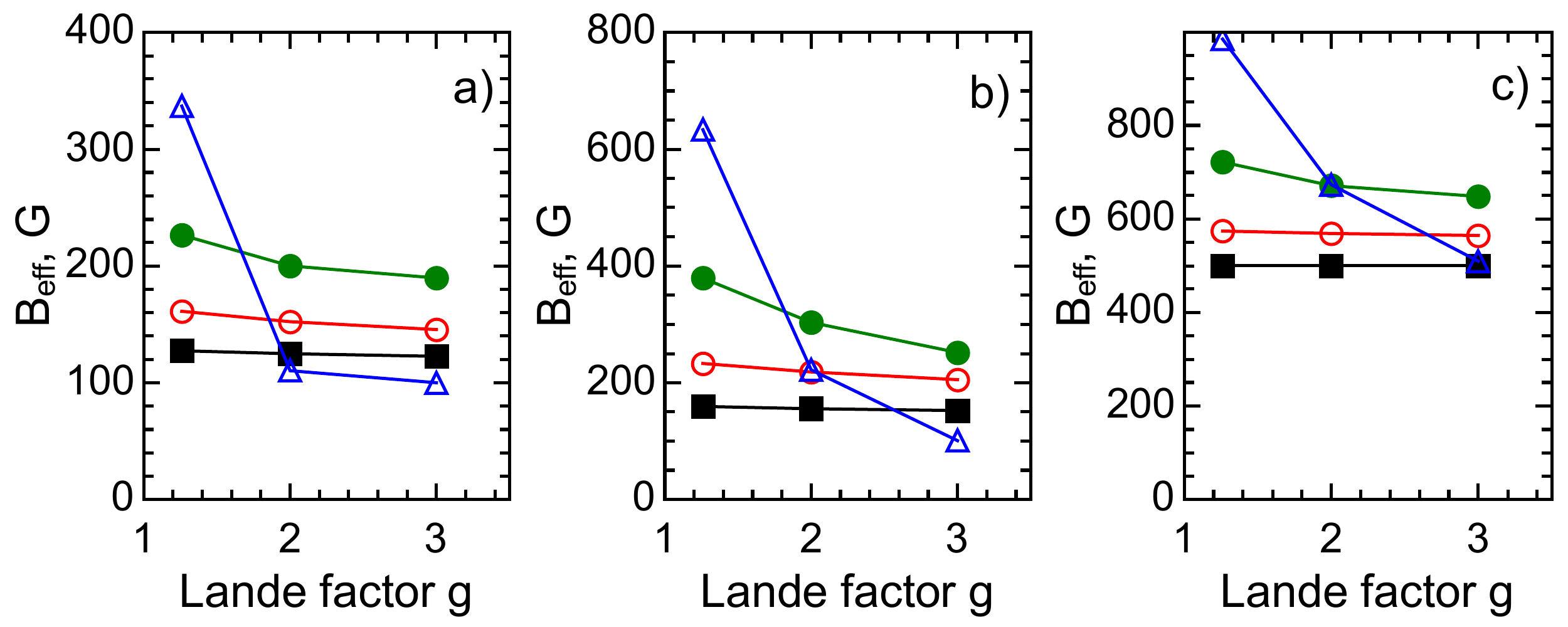}}
\caption{Effective magnetic field strengths $B_\mathrm{eff}$ as functions of Lande factor $g$ derived from synthetic line profiles of Fe~{\sc I} 5233~$\ang$ ($g=1.26$),
5247.1~$\ang$ ($g=2.00$) and 5250.2~$\ang$ ($g=3.00$) based on the two-component field model with the both, C1 and C2, components in absorption.  
Panel a is for the case with $B_\mathrm{back}=100$~G and $\Delta \lambda_{C2} = 0.25 \Delta \lambda_{C1}$; panel b is for the case with $B_\mathrm{back}=100$~G and $\Delta \lambda_{C2} = 0.5 \Delta \lambda_{C1}$; panel c is for the case with $B_\mathrm{back}=500$~G and $\Delta \lambda_{C2} = 0.5 \Delta \lambda_{C1}$
The field strength in C2 component is 500 G (black lines with solid squares), 1 kG (red with circles), 2 kG (green with solid circles), and 4 kG (blue with triangles).}
\label{f-babs}
\end{figure}

\begin{figure}[!h]    
\centerline{\includegraphics[width=1.0\textwidth,clip=]{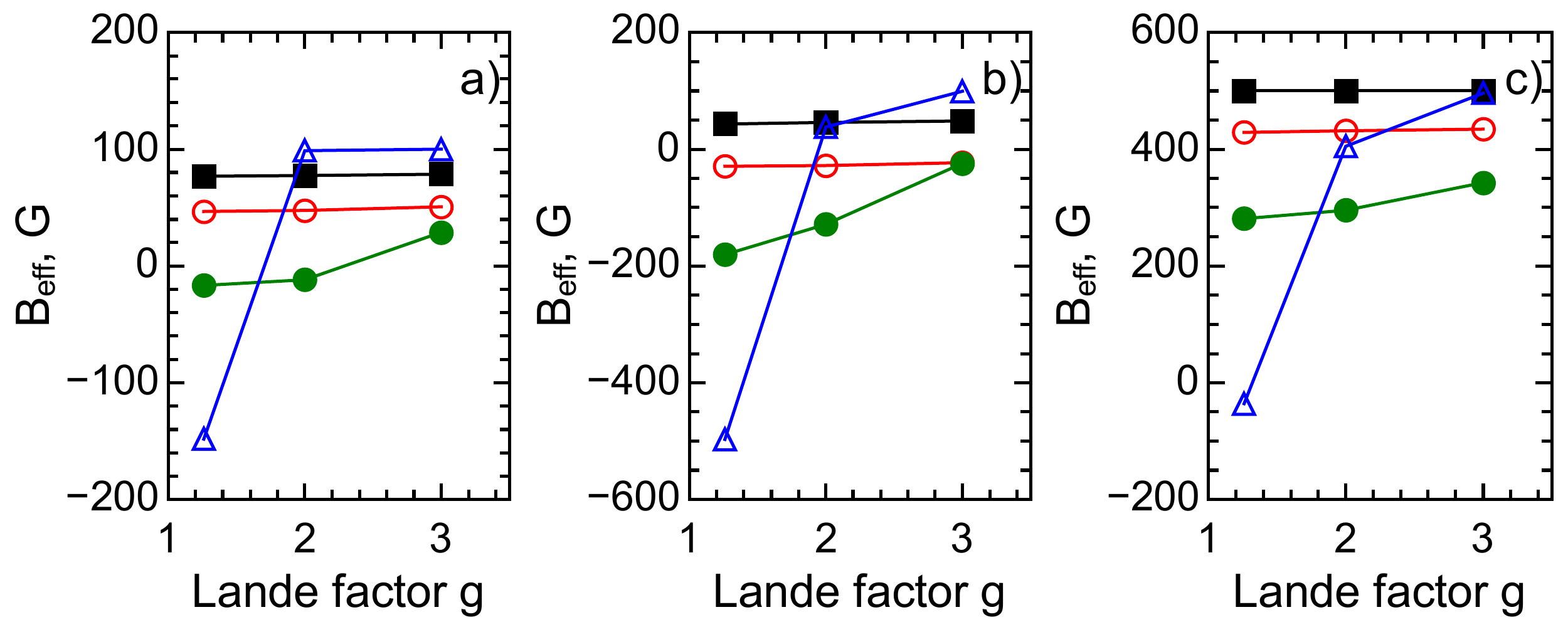}}
\caption{Effective magnetic field strengths $B_\mathrm{eff}$ as functions of Lande factor $g$ derived from synthetic line profiles of Fe~{\sc I} 5233~$\ang$ ($g=1.26$),
5247.1~$\ang$ ($g=2.00$) and 5250.2~$\ang$ ($g=3.00$) based on the two-component field model with C2 component in emission.  
Panel a is for the case with $B_\mathrm{back}=100$~G and $\Delta \lambda_{C2} = 0.25 \Delta \lambda_{C1}$; panel b is for the case with $B_\mathrm{back}=100$~G and $\Delta \lambda_{C2} = 0.5 \Delta \lambda_{C1}$; panel c is for the case with $B_\mathrm{back}=500$~G and $\Delta \lambda_{C2} = 0.5 \Delta \lambda_{C1}$
The field strength in C2 component is 500 G (black lines with solid squares), 1 kG (red with circles), 2 kG (green with solid circles), and 4 kG (blue with triangles).}
\label{f-bemi}
\end{figure}

Firstly, let us consider the classical saturation effect for the case when both spectral line components have absorption profiles.   Figure~\ref{f-saturation} shows the effective value of magnetic field depending on the Component 2 field strength for
constant background field anf filling factor. The effective magnetic field values appear to be between
$B_\mathrm{back}$ and $B_\mathrm{sub}$ when Component 2 yields absorption profiles and between $-B_\mathrm{sub}$ and $B_\mathrm{back}$ when Component 2 yields 
emission profiles. It can be seen that for small values of $B_\mathrm{sub}$ the effective field $B_\mathrm{eff}$ linearly increases 
(by absolute value) with $B_\mathrm{sub}$. However, when 

\begin{equation}\label{eq-saturation}
\Delta \lambda_{H\;  sub} \approx \Delta \lambda_{1/2}
\end{equation} 
then the contribution of the Component 2 to the total Zeeman splitting reduces with the increase of $\Delta \lambda_\mathrm{H\;  sub}$ and, hence, 
$B_\mathrm{eff}$ becomes nearly flat, and then starts to decrease (in absolute value) with $B_\mathrm{sub}$. Thus, according to 
Equation~\ref{eq-saturation}, the 5250.2~$\ang$ line has $\Delta \lambda_{1/2} \approx 65$~m$\ang$, and, hence, the function 
$B_\mathrm{eff}(B_\mathrm{sub})$ should saturate when $B_\mathrm{sub} \approx 0.9$~kG. For 5247.1~$\ang$ this field strength is 
$\approx 1.3$~kG, while
for 5233~$\ang$ line the saturation occurs only when the field in the Component 2 reaches $B_\mathrm{sub} \approx 5.5$~kG. Hence, when $B_\mathrm{sub} > 0.8 - 1.0$~kG
the effective field strength observed with the 5250~$\ang$ line is lower than that observed with the 5247.1~$\ang$ line, which, in turn, lower than the field
observed with the 5233~$\ang$ line. This effect is demonstrated by effective field values $B_\mathrm{eff}$ shown for different Lande factor values in Figures \ref{f-babs} and \ref{f-bemi}.

It can be seen that in the case of absorption in Component 2 (Figure~\ref{f-babs}) 
the $B_\mathrm{eff}(g)$ functions are nearly flat until $B_\mathrm{sub} \approx 1.0$~kG. 
For higher values of $B_\mathrm{sub}$ the 
$B_\mathrm{eff}(g)$ functions have negative inclinations. The ratio $B_\mathrm{eff\;5233}/B_\mathrm{eff\;5250}$ shows little dependence on the width 
of the Component 2 profile but depends strongly on the background field strength: low $B_\mathrm{back}$ result in high $B_\mathrm{eff\;5233}/B_\mathrm{eff\;5250}$.

In the case of Component 2 emission (Figure~\ref{f-bemi}) the picture is opposite. Similar to the absorption case, 
the $B_\mathrm{eff}$ values are nearly equal while $B_\mathrm{sub}$
is below $1-1.5$ kG. However, when 5250.2~$\ang$ and 5247.1~$\ang$ lines start to saturate, the $B_\mathrm{eff}(g)$ functions have positive inclination. When the
background field is low ($B_\mathrm{back} \approx 100$~G, Figs.~\ref{f-bemi}(a-b)), the splitting of 5233~$\ang$ line is dominated by Component 2 and the values 
of $B_\mathrm{eff}$ can be substantially negative. However, in the case of higher background field ($0.5$~kG, Figure~\ref{f-bemi}(c)) the $B_\mathrm{eff}$ have the
same sign in all lines, yielding the ratios $0 < B_\mathrm{eff\;5233}/B_\mathrm{eff\;5250} < 1$, similar to those observed in flares.


Obviously, the magnetic field value, at which saturation starts, also depends on how $B_\mathrm{eff}$ is calculated. The 
centres-of-mass of $I+V$ and $I-V$ Stokes profiles are calculated using the areas outlined by the profiles and a certain intensity
level $i_\mathrm{level}$. In the present study, $i_\mathrm{level}$ is set at a half-depth on an observed $I\pm V$ profile. Setting this level at lower or higher intensity would lead to the saturation at lower or higher $B$, respectively. This is because lowering $i_\mathrm{level}$ effectively means limiting maximum value of magnetic field that can contribute to $B_\mathrm{eff}$. However, although it would be logical to increase the value of $i_\mathrm{level}$, it can lead to very substantial errors in $B_\mathrm{eff}$ measurements: normally, line profiles above $\approx 0.7-0.8$ of their depth become noisy and affected by blends.

\subsection{Bisector Splitting Functions Derived from the Synthetic Profiles}\label{synt-bs}

\begin{figure}[!hb]   
\centerline{\includegraphics[width=1.0\textwidth,clip=]{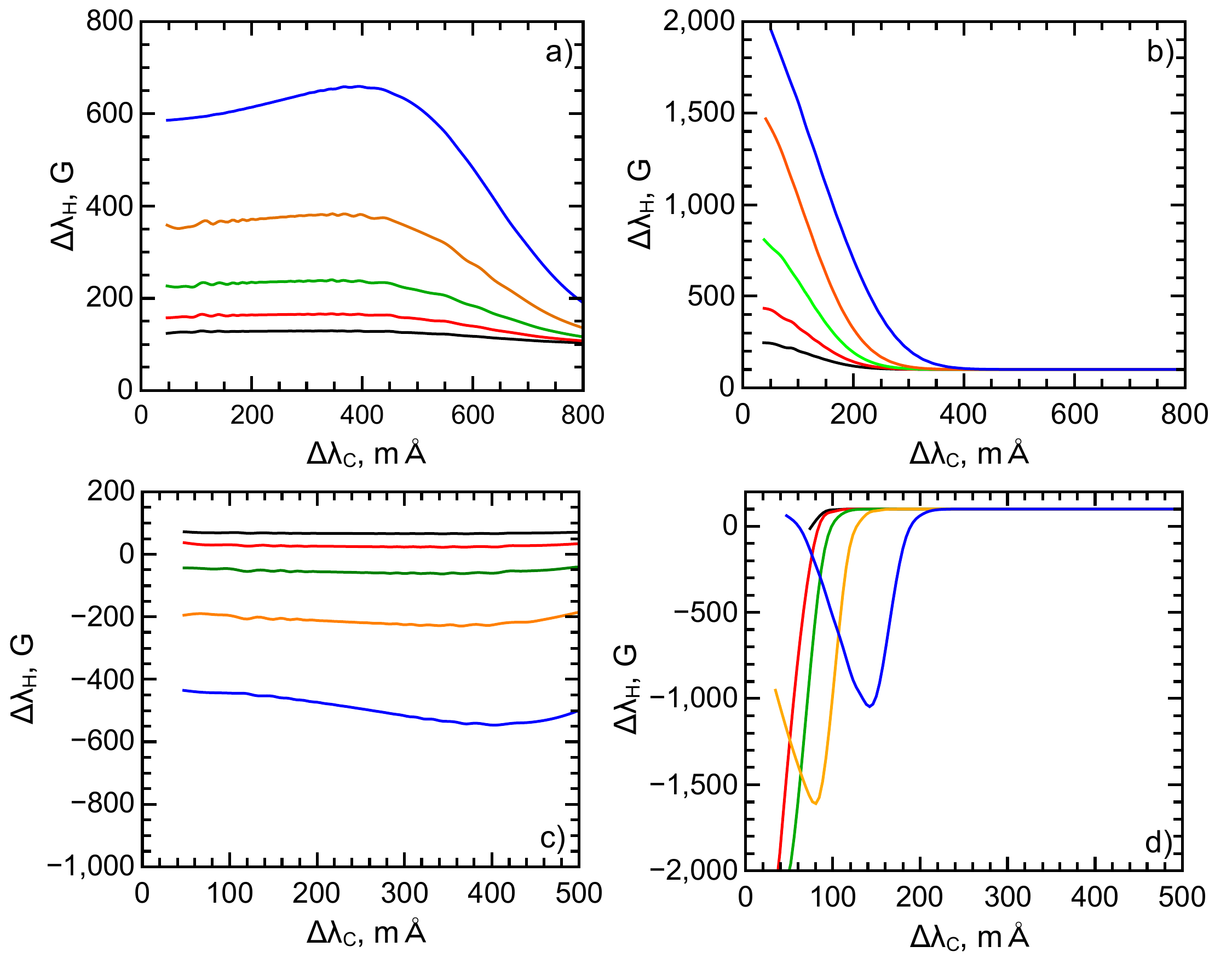}}
\caption{Bisector splitting functions for synthetic 5233~$\ang$ $I\pm V$ profiles. Panels a and b are for cases when C2 component has absorption profile, panels c and d are for cases when C2 component has emission profile. Panels (a) and (c) are for cases when C2 and C1 components have the same widths, while panels (b) and (d) are for cases when the widths of C2 components are three times smaller than the widths of C1 components. The background field is 100 G in all the cases. Black lines are for $B_\mathrm{sub} = 250$~G, red lines -- 500~G, green lines -- 1~kG, orange lines -- 2~kG, blue lines -- 4~kG.}
\label{f-syntbis}
\end{figure}

Bisector splitting functions $\Delta \lambda_\mathrm{H}(\Delta \lambda_\mathrm{c})$ derived from the synthetic $I\pm V$ profiles are shown in Figure~\ref{f-syntbis}. It can be seen that the bisector splitting 
values always remain between $B_\mathrm{back}$ and  $B_\mathrm{sub}$ in case of C2 absorption and between $-B_\mathrm{sub}$ and $B_\mathrm{back}$ in case of C2 emission. The splitting 
is higher in the line core and then drops towards the wings (similar to \citext{ulre09}). If the width of the Component 2 profile is similar to that of the background C1 profile, the 
bisector splitting function is very smooth, slowly decreasing to the $B_\mathrm{back}$ value. However, if the Component 2 profile is much narrower, it results
in very strong inclination of the bisecor splitting function when $B_\mathrm{sub}$ is relatively low (lower than $\approx 5$~kG), or in appearance of localised deviations
when $B_\mathrm{sub}$ is high. Namely, bisector splitting functions $\Delta \lambda_\mathrm{H}(\Delta \lambda_\mathrm{c})$ will show a relatively narrow peak in case of C2 absorption and a relatively narrow deep in case of C2 emission. Doppler shifts of C2 component should not affect
the effective field values $B_\mathrm{eff}$ and bisector splitting functions, provided C1 and C2 profiles are rather similar and the magnetic field in C2 component is low ({\it i.e.} below the saturation value. Otherwise, unresolved velocity field inhomogeneity can 
affect magnetic field measurements.

This behaviour makes it possible to explain the appearance of localised variation in the bisector splitting functions observed in flares: the peaks result from the presence of the narrow C2 component (either in absorption or emission). 

\subsection{The effect of Doppler shift on the effective magnetic field values and bisector splitting}\label{synt-doppler}

\begin{figure}    
\centerline{\includegraphics[width=0.55\textwidth,clip=]{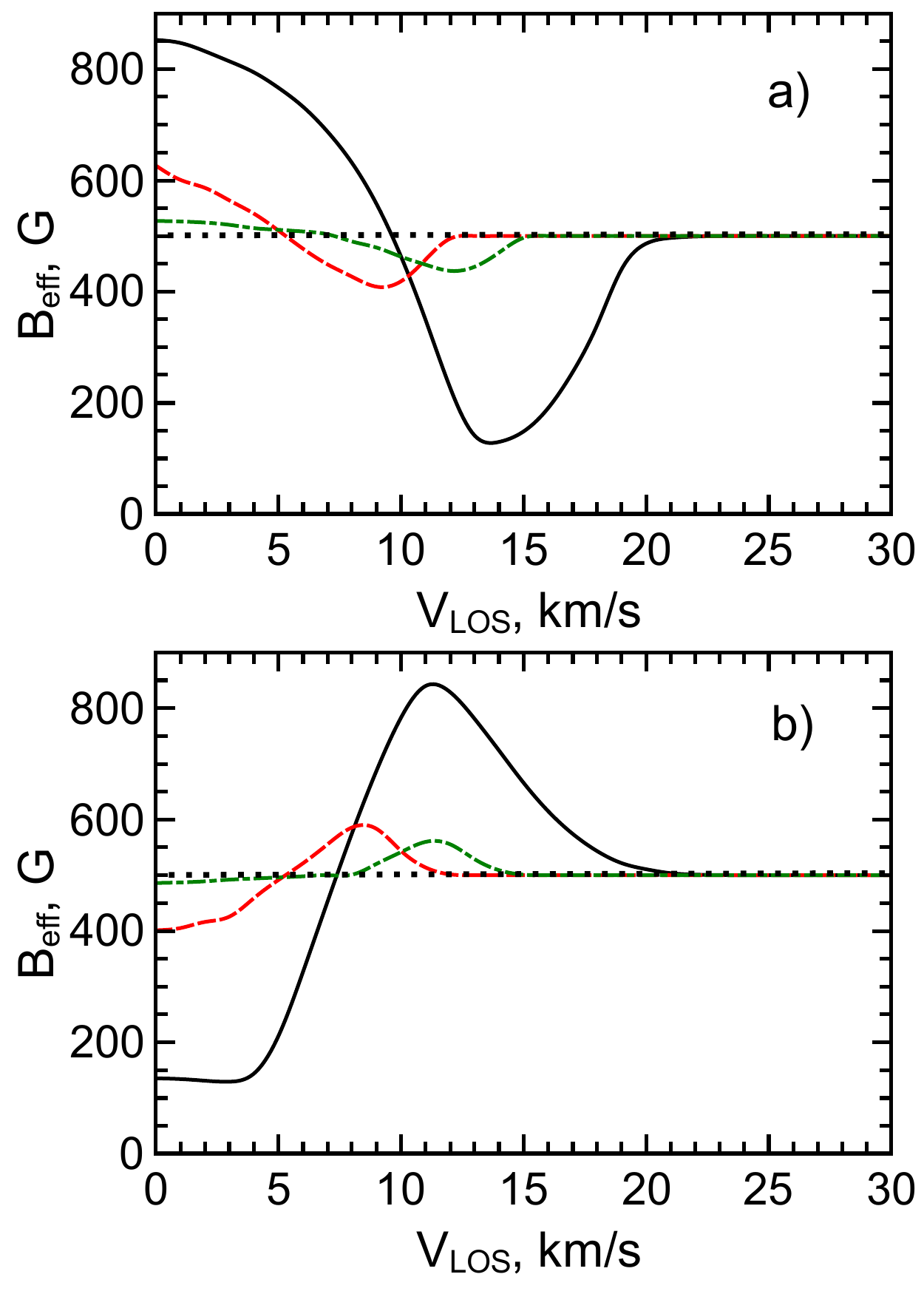}}
\caption{Variation of the effective magnetic field strength due to the Doppler shift of the C2 component (in respect of C1 component) in the synthetic profiles of Fe~{\sc I} 5233~$\ang$ (black solid lines), 5247.1~$\ang$ (red dashed lines), and 5250.2~$\ang$ (green dot-dashed lines). The background field is 500 G and the field strength in C2 component is 4 kG; 
the filling factor is 10\,\%; C2 profiles have widths twice smaller than the width of the C1 profile. Panel a is for C2 in absorption; panel b is for C2 in emission.}
\label{f-doppler}
\end{figure}

So far we assumed that the Doppler shift of Component 2 profiles relative to Component 1 profiles is zero. However, in many flares  emission peaks are visibly blue- or red-shifted relative to the absorption profile. Even without visible emission peaks, $I\pm V$ profiles often demonstrate asymmetry, giving strong indication of the Doppler effect.

The effect of Doppler shift of Component 2 on $B_\mathrm{eff}$ measurements is shown in Figure~\ref{f-doppler}. In the case of absorption, non-zero LOS velocity in unresolved flux tubes generally results in reduction of the C2 contribution to the overall Zeeman splitting and, hence, lower effective field magnitude $B_\mathrm{eff}$. Thus, when the Doppler shift is relatively low
($\Delta \lambda_\mathrm{LOS} \lesssim \Delta \lambda_{1/2}$) the value of $B_\mathrm{eff}$ is lower than that in the case $\Delta \lambda_\mathrm{LOS}=0$, but remains higher than $B_\mathrm{back}$. At higher Doppler shifts 
($\Delta \lambda_{1/2} \lesssim \Delta \lambda_\mathrm{LOS} \lesssim 2 \Delta \lambda_{1/2}$) the effective field strength drops below the
background field strength. Once the Doppler shift is noticeably larger than $2 \Delta \lambda_{1/2}$,  the contribution of
the Component 2 to the resulting $I\pm V$ profiles becomes negligible, and the effective field becomes equal to $B_\mathrm{back}$. 

In the case of emission, the picture is similar, although the change in effective field value is opposite: the value of 
$B_\mathrm{eff}$ increases with $\Delta \lambda_\mathrm{LOS}$  and becomes equal $B_\mathrm{back}$ when 
$\Delta \lambda_\mathrm{LOS} \approx \Delta \lambda_{1/2}$. When 
$\Delta \lambda_{1/2} \lesssim \Delta \lambda_\mathrm{LOS} \lesssim 2 \Delta \lambda_{1/2}$  
the effective field strength is above $B_\mathrm{back}$, and then drops to $B_\mathrm{back}$ when 
$\Delta \lambda_\mathrm{LOS} \gtrsim 2 \Delta \lambda_{1/2}$.

\subsection{Comparison of Synthetic and Observed Profiles for Three Flares}\label{synt-flares}

\begin{figure}    
\centerline{\includegraphics[width=0.75\textwidth,clip=]{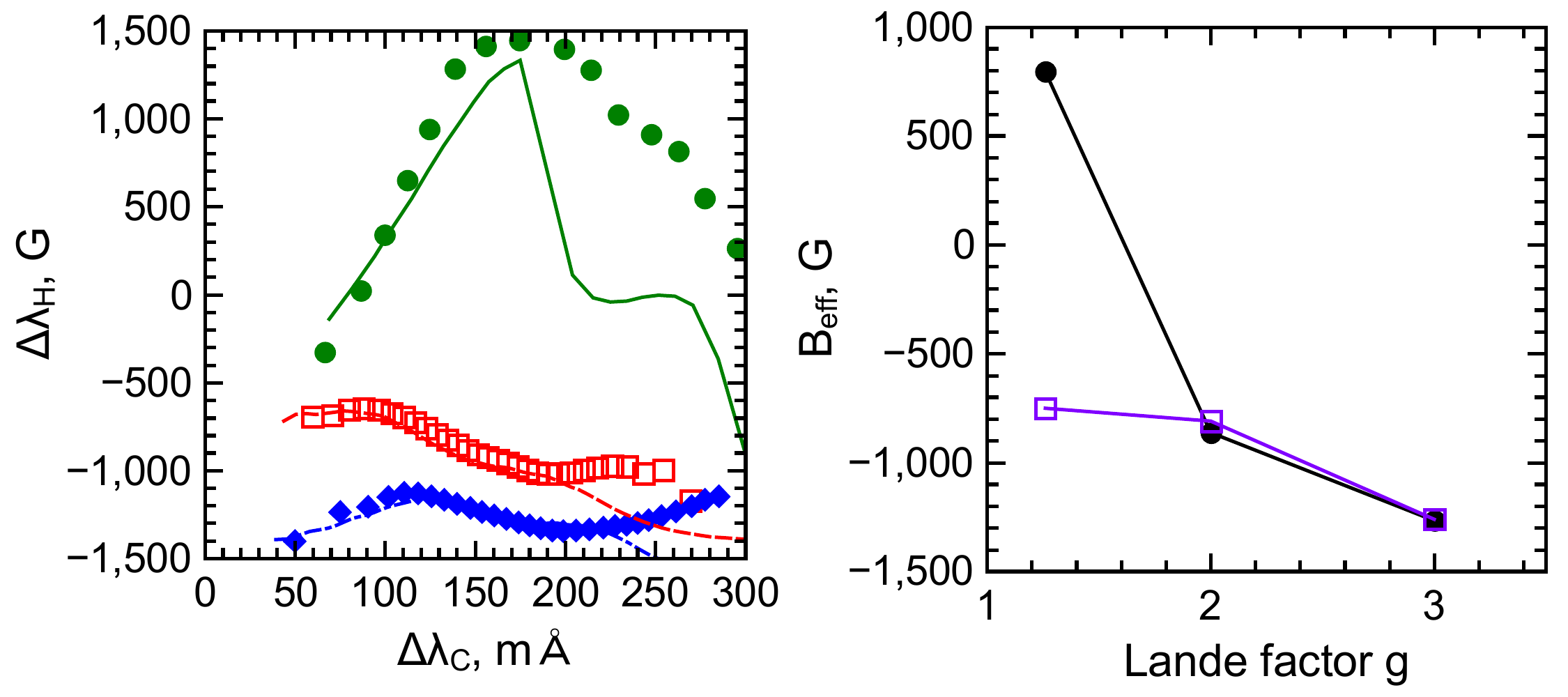}}
\caption{Comparison of the observed and synthetic $I\pm V$ profiles of Fe~{\sc I} 5233~$\ang$, 5247.1~$\ang$, and 5250.2~$\ang$ lines in the flare 5.
Left panel: Bisector splitting functions (solid green line and solid circles -- 5233~$\ang$, dashed red line and squares -- 5247.1~$\ang$, dot-dashed blue line and triangles -- 5250.2$\ang$; observed and synthetic are shown as symbols and lines, respectively). 
Right panel: Effective magnetic field values deduced from observed (black line with circles) and synthetic (purple line with squares) profiles.}
\label{f-bis5}
\end{figure}

\begin{figure}    
\centerline{\includegraphics[width=0.75\textwidth,clip=]{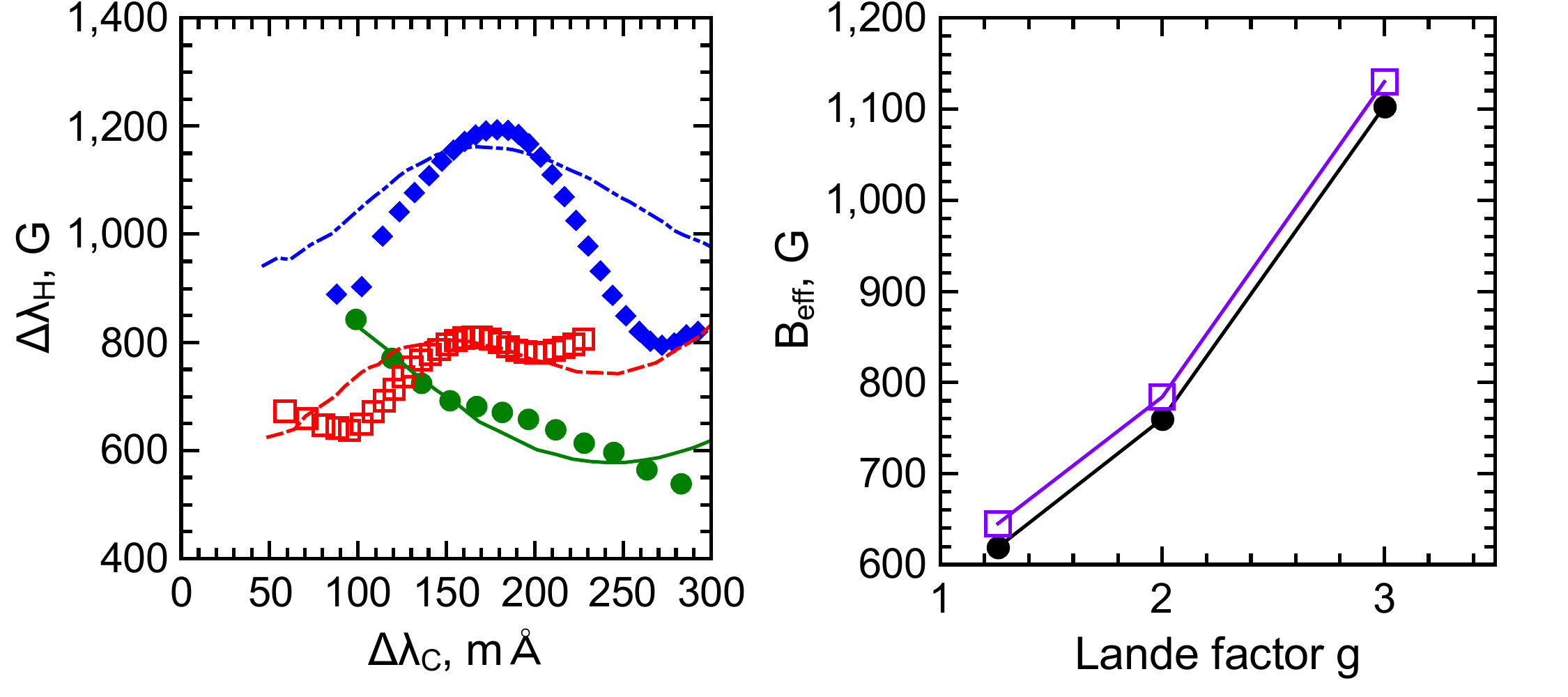}}
\caption{Comparison of the observed and synthetic $I\pm V$ profiles of Fe~{\sc I} 5233~$\ang$, 5247.1~$\ang$, and 5250.2~$\ang$ lines in the flare 6.
Left panel: Bisector splitting functions (solid green line and solid circles -- 5233~$\ang$, dashed red line and squares -- 5247.1~$\ang$, dot-dashed blue line and triangles -- 5250.2$\ang$; observed and synthetic are shown as symbols and lines, respectively). 
Right panel: Effective magnetic field values deduced from observed (black line with circles) and synthetic (purple line with squares) profiles.}
\label{f-bis6}
\end{figure}

\begin{figure}    
\centerline{\includegraphics[width=0.75\textwidth,clip=]{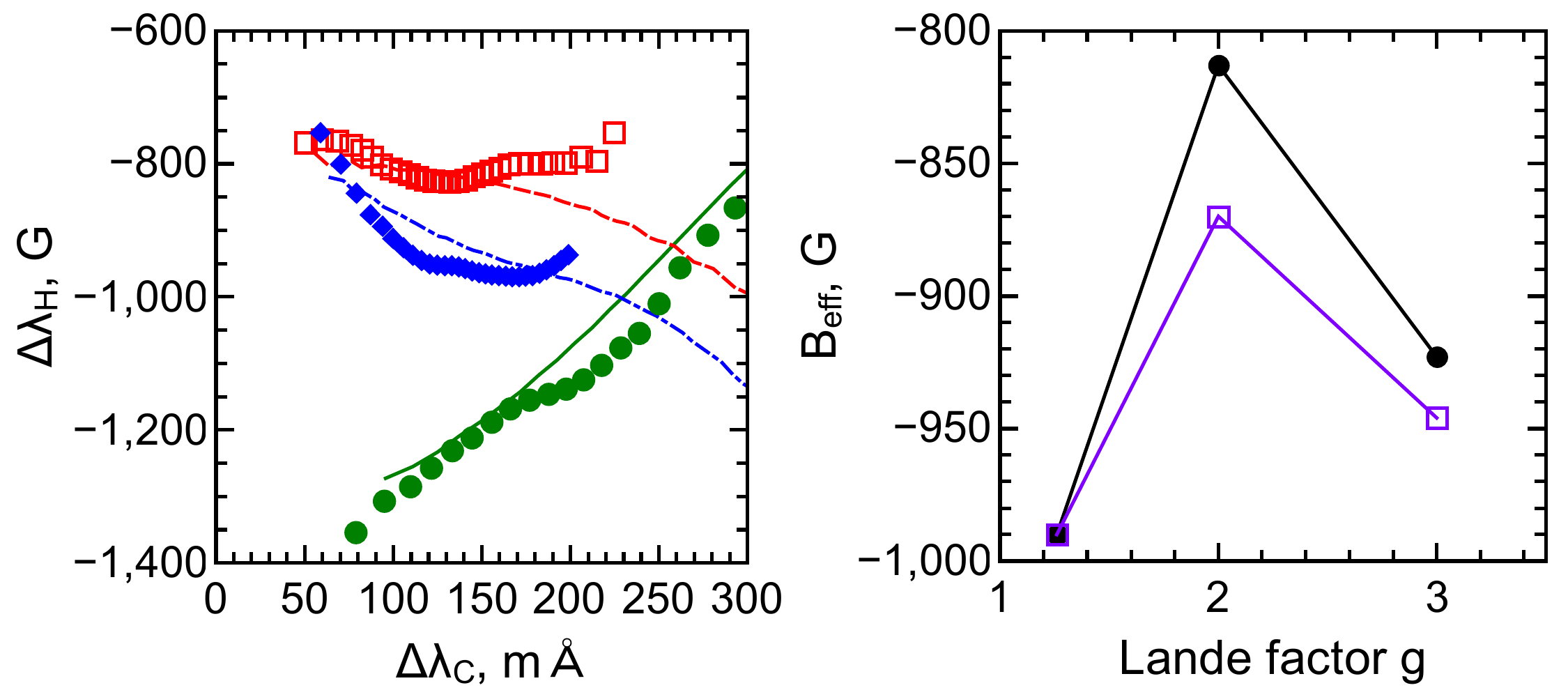}}
\caption{Comparison of the observed and synthetic $I\pm V$ profiles of Fe~{\sc I} 5233~$\ang$, 5247.1~$\ang$, and 5250.2~$\ang$ lines in the flare 11.
Left panel: Bisector splitting functions (solid green line and solid circles -- 5233~$\ang$, dashed red line and squares -- 5247.1~$\ang$, dot-dashed blue line and triangles -- 5250.2$\ang$; observed and synthetic are shown as symbols and lines, respectively). 
Right panel: Effective magnetic field values deduced from observed (black line with circles) and synthetic (purple line with squares) profiles.}
\label{f-bis11}
\end{figure}

In this section we attempt to deduce C1 and C2 parameters by fitting observed line profiles with the synthetic profiles based on the two-component field model. We choose three flares, 5, 6, and 11, which have relatively low noise level in their Stokes profiles and demonstrate rather ``well-behaved'' bisectors. Figures \ref{f-bis5}, \ref{f-bis6}, and \ref{f-bis11} compare bisector splitting functions and effective field strengths for three spectral lines in these flares. 

In order to fit the observational data, we use the following procedure: first, we evaluate the background field using $I\pm V$
splitting of the observed profile wings. Then we vary and adjust the filling factor, $\Delta \lambda_\mathrm{H}$, $\Delta \lambda_\mathrm{LOS}$, and $\Delta \lambda_{1/2}$ in order to minimise the deviation between the bisector splitting functions of the observed and synthetic profiles. 
The fitting parameters for background and strong component are given in Table~\ref{table-fit}.

The parameters of the magnetic field components derived from different spectral lines appear to be quite close in flares 6
and 11: deviations in magnetic field strengths $B_\mathrm{back}$, $B_\mathrm{sub}$, and Doppler velocities ($\Delta \lambda_D$)  are below 25\,\%. The field values $B_\mathrm{back}$ used to fit observed 5247.1~$\ang$ and 5250.2~$\ang$ profiles are systematically higher (although by no more than 20\,\%) than those for 5233~$\ang$ line. This can be easily explained by vertical gradient of magnetic field, as 5233~$\ang$ line is formed slightly higher than 5247.1~$\ang$ and 5250.2~$\ang$ lines.

In the flare 5 the $B_\mathrm{back}$ and $\mathcal{F}$ value are in quite good agreement, while the $B_\mathrm{sub}$ value required for the 5233~$\ang$ line (5.5 kG),
which is much higher that $B_\mathrm{sub}$ for 5247.1~$\ang$ and 5250.2~$\ang$ lines ($\approx$ 2.5 kG) and, in general, is much higher than any previous field measured or
estimated outside sunspots. This flare requires further investigation.

In all these three flares the temperatures (or C2 profile widths) deduced from these three lines are quite different (sometimes by factor of up to two). This discrepancy
is likely to include two effects: actual profile width difference due to different temperature sensitivity and, inevitably, a fitting error.

\begin{table}[!ht]
\caption{Magnetic field characteristics derived for three solar flares (two X-class flares that occured on 02 Apr. 2001, flares 5 and 6, and one C-class
flare that occured on 10 May 2012, flare 11). The values in the fourth column are product of the filling factor (as per Equation~\ref{eq-combi}) and the amplitude 
of the synthtic emission or absorption profile (negative values correspond to emission, positive values to absorption). The sixth column gives temperature 
corresponding to the doppler width of the synthtic emission or absorption components corresponding to $B_\mathrm{sub}$.}
\label{table-fit}
\begin{tabular}{|l|l|l|l|l|l|}
\hline
Flare 5 & $B_\mathrm{back}$, G & $B_\mathrm{sub}$, G & $\mathcal{F}$ & $\Delta \lambda_\mathrm{LOS}$, km s$^{-1}$ & $T_\mathrm{sub}$, 10$^3$~K \\
\hline
5233.0~$\ang$ & -700 & -5500 & -0.09 & -20 & 10 \\
5247.1~$\ang$ & -700 & -2700 & -0.09 & -20 & 10 \\
5250.2~$\ang$ & -890 & -2500 & -0.08 & -18 & 8 \\
\hline      
\hline
Flare 6 & & & & & \\
\hline
5233.0~$\ang$ & 480 & 3000 & -0.04 & -60 & 23 \\
5247.1~$\ang$ & 480 & 3000 & -0.04 & -40 & 11 \\
5250.2~$\ang$ & 600 & 2750 & -0.05 & -18 & 10 \\
\hline      
\hline
Flare 11 & & & & & \\
\hline
5233.0~$\ang$ & -680 & -1500 & -0.12 & -50 & 30 \\
5247.1~$\ang$ & -650 & -1750 & -0.12 & -25 & 20 \\
5250.2~$\ang$ & -670 & -1750 & -0.12 & -30 & 18 \\
\hline
\end{tabular}
\end{table}

\subsection{Estimated Error of Bisector Splitting Measurements Based on the Synthetic Profiles}\label{synt-error}

\begin{figure}    
\centerline{\includegraphics[width=1.0\textwidth,clip=]{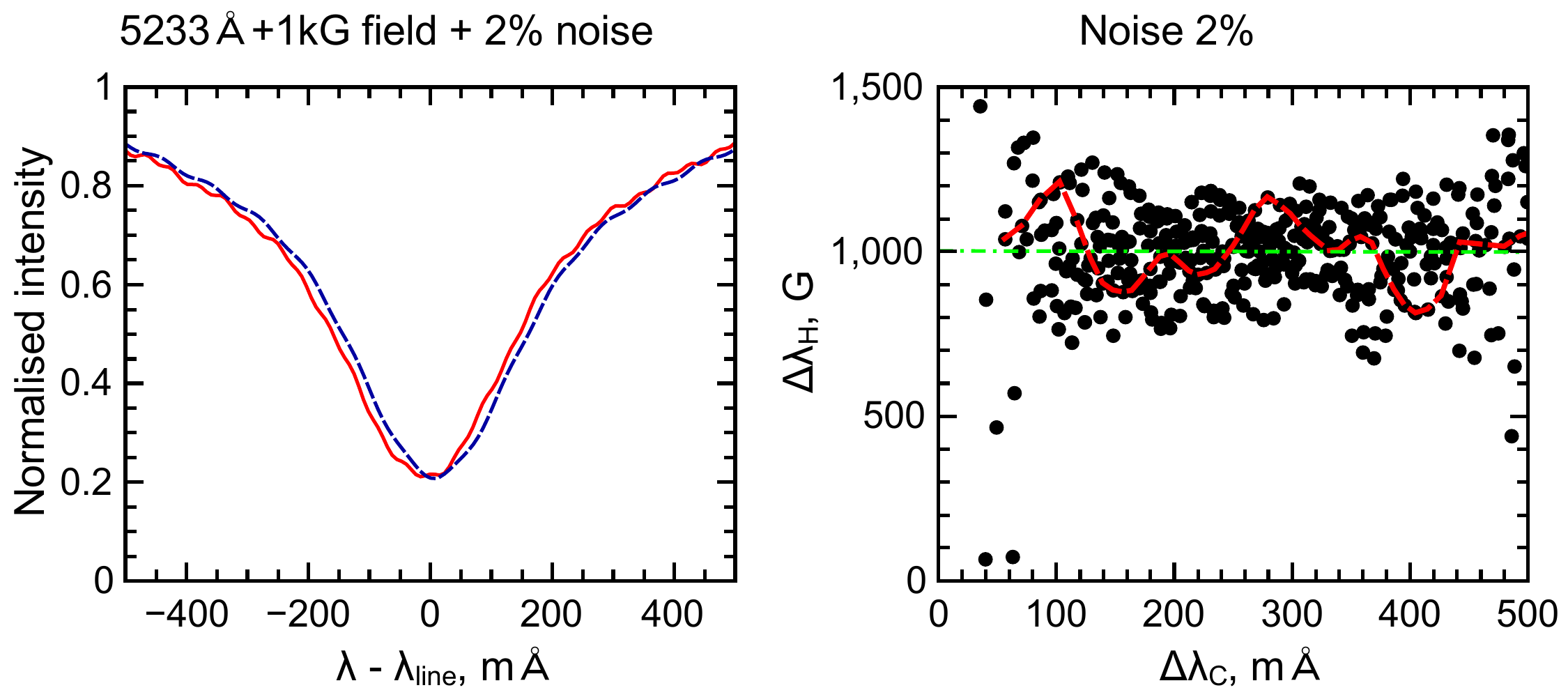}}
\caption{The effect of photometric noise on the bisector splitting functions. Left panel shows synthetic $I\pm V$ profiles of Fe~{\sc I} 5233~$\ang$ line 
calculated for uniform magnetic field of 1 kG with added 2\,\% noise. Right panel shows scatter plots demonstrating deviations of measured 
$\Delta \lambda_\mathrm{H}(\Delta \lambda_\mathrm{c})$ values from the exact value (shown as green dot-dashed line) for 2\,\% noise. Red dashed line in the right panel is the bisector splitting function corresponding to the profile shown in the left panel.}
\label{f-biserror}
\end{figure}

\begin{figure}   
\centerline{\includegraphics[width=0.55\textwidth,clip=]{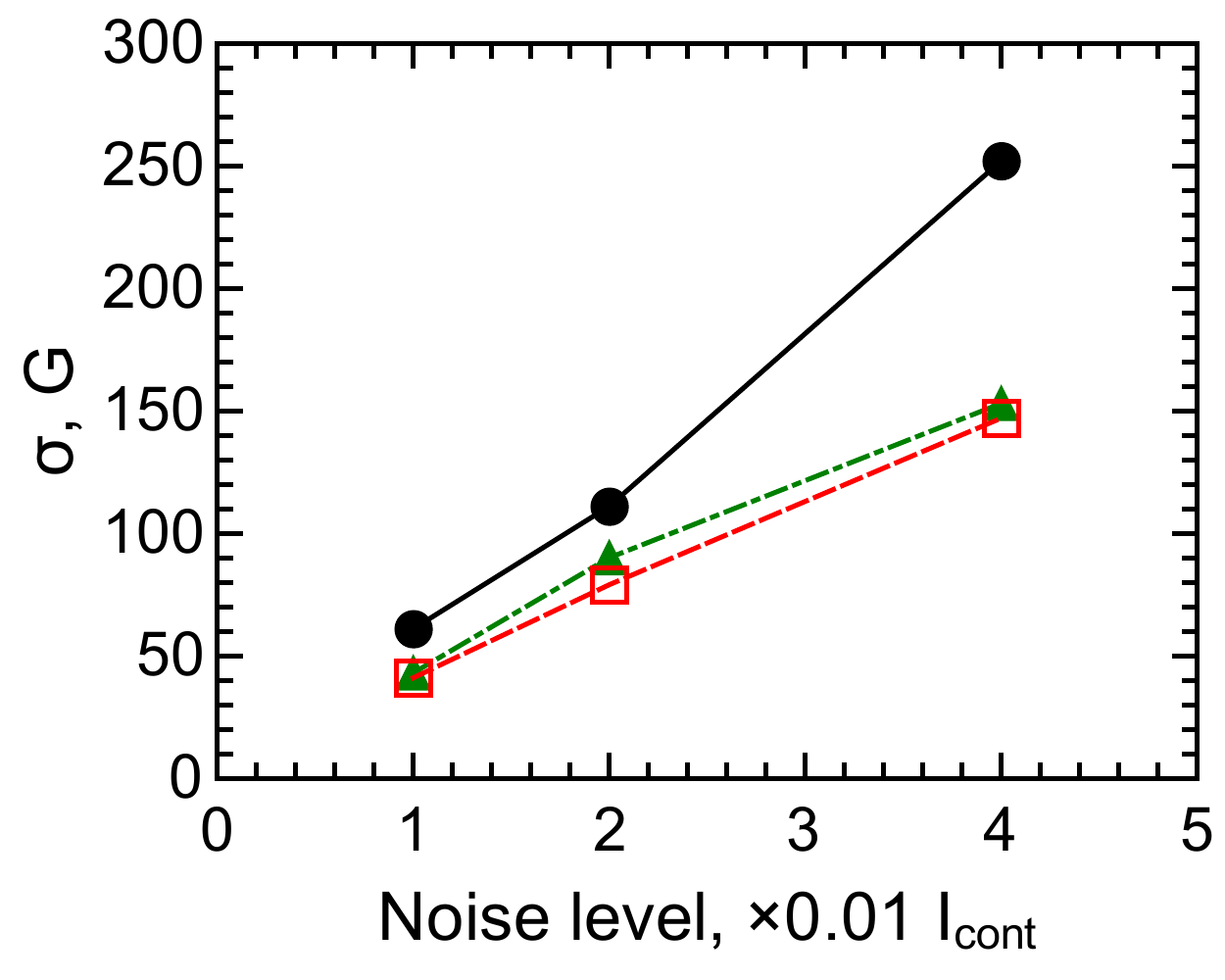}}
\caption{Mean square error of the effective field strength measurements versus noise level for the 5233~$\ang$ (solid black line, solid circles), 5247.1~$\ang$ (dashed red line with open squares), and 5250.2~$\ang$ (dot-dashed green line with triangles) spectral lines.}
\label{f-raterror}
\end{figure}

Synthetic line profiles allow us to evaluate the typical error in Zeeman effect measurements. Figure~\ref{f-biserror} demonstrates the error in bisector splitting
values in 5233~$\ang$ line, while Figure~\ref{f-raterror} shows the typical error in $B_\mathrm{eff}$ measurements. 

It can be seen, that the error in bisector splitting measurements in the range $\Delta \lambda_\mathrm{c}= 100 - 300$~m$\ang$ is 
$\approx 70-100$~G when the noise amplitude is 1\,\% (of the continuum intensity); this error is $\approx 150$~G and $250 - 260$~G when the noise amplitudes are
2\,\% and 4\,\%, respectively. In the observations presented in Section ~\ref{observ}, typical noise level is around 2\,\% and, hence, the error is $\approx 150$~G, which
is close to that evaluated from the telluric line observations -- $\approx 100$~G. It should be noted, that the error is higher at the line core 
($\Delta \lambda_\mathrm{c}< 100$~m$\ang$ for the 5233~$\ang$ line) and close to the wings ({\it i.e.} at large $\Delta \lambda_\mathrm{c}$). This is due to the fact that the error is higher for lower 
$dI/d\lambda$ values, given the same noise amplitude. This also means that measurements in narrow lines, such as 52471.$\ang$ and 5250.2~$\ang$, have substantially lower error.

$B_\mathrm{eff}$ values are much less sensitive to the noise. Thus, the noise with an amplitude of 1\,\% yields mean square error of $\approx 3$~G for 
5247.1~$\ang$ and 5250.2~$\ang$ lines and $\approx 6$~G forthe  5233~$\ang$ line. For noise amplitudes of 2\,\% these values are $\approx 7$~G and $\approx 10$~G, and for the 
4\,\% noise these errors reach $\approx 14$~G and $\approx 25$~G, respectively. Hence, the comparison of $B_\mathrm{eff}$ field values deduced from spectral lines with different magnetic sensitivities would provide more reliable information regarding unresolved magnetic field structure than bisector splitting functions.

\section{Discussion}
	\label{concl}

Our observational data demonstrates that, similarly to the quiet photopshere \cite{sten73,sola93}, the magnetic field at 
the photospheric level in flares is very inhomogeneous at unresolved scales. The observational picture is more complicated
than in the quiet photosphere and in plages due to the presence of emission and fast plasma flows along the line-of-sight. 

The comparison of the magnetic field values deduced from spectral lines with different Lande factor $g$ shows that the
effective field strength $B_\mathrm{eff}$ increases with $g$ (see Section ~\ref{observ-lr}), in contrast with what is normally 
observed in the quiet photosphere. Analysis of the synthetic $I\pm V$ Stokes profiles for two-component fields shows that 
this is possible in two cases: when the strong magnetic field has its polarity opposite to the polarity of the weaker 
ambient field or when the spectral components corresponding to the unresolved field (C2) show emission. The presence of very
strong field of opposite polarity should be associated with high current densities. Hence, the second possibility, with
emission from intense magnetic flux tubes, looks more realistic. Furthermore, the second option seems to be more viable as 
emission peaks are often observed in metallic lines in moderate and bright flares. 

The fine structure of $I\pm V$ profiles observed in flares has been studied using bisector splitting functions ($\Delta \lambda_\mathrm{H}(\Delta \lambda_\mathrm{c})$) 
(Section ~\ref{observ-bs}). 
It is known that the centre of a Fraunhofer line is formed predominantly in cooler regions, while the wings 
correspond to higher temperatures and are formed slightly deeper. Hence, the average trend of 
$\Delta \lambda_\mathrm{H}(\Delta \lambda_\mathrm{c})$ can be considered
as the result of corellation between the magnetic field and the temperature. Hence, it is very unlikely, that the decrease of $\Delta \lambda_\mathrm{H}$ with 
$\Delta \lambda_\mathrm{c}$ is the result of vertical magnetic field gradients, as it would imply magnetic field increase with height. Horizontal inhomogeneity of the magnetic field  
seems to be rather realistic explanation for the observed trends of $\Delta \lambda_\mathrm{H}(\Delta \lambda_\mathrm{c})$, with higher magnetic field
corresponding to lower temperatures and turbulent velocities. Analysis of the bisector splitting in synthetic $I\pm V$ Stokes profiles shows that the localized extrema, similar 
to those seen in Figure~\ref{f-bsall}, can be explained by the presence of narrow emission or absorption components of the 
spectral line with substantial shift $\Delta \lambda_\mathrm{H}$. Thus, based on synthetic profiles, it is possible to 
estimate the magnitude of the field, that would result in peaks on the bisector splitting profiles. The peak at $\Delta \lambda_\mathrm{c}\approx 75$~m$\ang$ in the case of the line similar to Fe~{\sc I}~$5233$~$\ang$ can be caused by
field strength $\approx 2.5-3.0$~kG. However, these estimations are very sensitive to the width of the main component of
the spectral line, and should be used with caution.

In general, our preliminary results confirm previous conclusions that the photospheric magnetic field in flares has at least two components. Furthermore, our study shows that in flares thermodynamic conditions in the intense magnetic flux tubes are
very different from conditions outside. The most interesting feature 
that has been revealed by this study, is an apparent link between the strong unresolved field and emission in flares. This
effect can be easily seen in observed profiles: the Zeeman split of emission peaks in line cores is normally bigger than that of the absorption component of a spectral line. This finding could be quite important for the flare phenomenology. There are several 
possible explanations for emission observed in photospheric lines: atoms can be excited due to non-thermal particle precipitation, heating by propagating waves, or by conduction. In any case, it is very likely that the observed emission is directly related to energy release in the corona. Hence, the revealed connection between the emission in metallic lines and stronger field component may indicate that the intense unresolved photospheric 
magnetic elements are topologically connected to the coronal field, while weak ambient photospheric magnetic fluxes only form the
low-level magnetic canopy (see, {\it e.g.}, \opencite{sole99}).

This study gives a rough estimate of the backgound and strong photospheric magnetic field components. Obviously, more work
needs to be done before these can be evaluated more reliably. Observationally, higher precision may be achieved by using
combination of different methods, for instance, by using the data obtained from Zeeman and Hanle measurements.
This, however, would not help to answer the question about the size of the small-scale magnetic elements. In order to address 
this problem, direct observations with higher spatial resolution are needed and these are likely to be possible with future 
missions. Current instruments provide reasonably high spatial resolution of $\approx 200-300$ km, but this is still not sufficient 
for reliable investigation of magnetic field fine structure. Thus, high-resolution magnetic field maps of flares 12 and 13 
obtained with Helioseismic and Magnetic Imager (HMI) onboard Solar Dynamic Observatory (SDO) show that the field is still very inhomogeneous even on a scale of one pixel.
  
Alternatively, observations in other spectral bands may provide
an opportunity to resolve very small scales. For example, future solar observations with Atacama Large Millimeter/submillimeter Array (ALMA) would be able to provide polarimetric data in sub-THz range with mili-arcsec resolution, possibly giving a unique insight into solar magnetic field structure \cite{loue09}.
Additionally, local helioseismology may provide an indirect estimations, as $p$-mode scattering and absorption can be sensitive to the magnetic flux tube sizes (see, {\it e.g.}, \opencite{choe98}; \opencite{goja08}; \opencite{jago08}; \opencite{fele12}). Finally, more theoretical work concerning the radiative transfer in strongly inhomogeneous magnetic field is needed in order to explain adequately the observed fine structure of Stokes profiles.

\begin{acks}
The authors thank Philippa Browning for useful comments from which the article strongly benefited. 
MG is supported by the Science and Technology Facilities Council (UK).
\end{acks}

\end{article} 

\end{document}